\documentclass[aps,prl,twocolumn,showpacs,superscriptaddress,tightenlines,preprintnumbers,amsmath,amssymb,floatfix,nofootinbib]{revtex4-1}
\usepackage{graphicx}
\usepackage{color}

\usepackage{etoolbox}
\let\bbordermatrix\bordermatrix
\patchcmd{\bbordermatrix}{8.75}{4.75}{}{}
\patchcmd{\bbordermatrix}{\left(}{\left[}{}{}
\patchcmd{\bbordermatrix}{\right)}{\right]}{}{}

\def\extra#1{{``#1''}}

\def \etal{\textit{et al.}}

\newcommand{\sr}{Sci. Rep.}

\def\tr{{\rm Tr}}

\def\<{\langle}
\def\>{\rangle}

\newcommand{\ket}[1]{\mbox{$|#1\rangle$}}
\newcommand{\bra}[1]{\mbox{$\langle#1|$}}
\newcommand{\oprod}[1]{\mbox{$|#1\rangle\langle#1|$}}

\def \info#1{}

\begin{document}



\title{Priority Choice Experimental Two-qubit Tomography: \\ Measuring
One by One All Elements of Density Matrices}

\author{Karol Bartkiewicz}
\email{bark@amu.edu.pl} \affiliation{Faculty of Physics, Adam
Mickiewicz University, PL-61-614 Pozna\'n, Poland}
\affiliation{RCPTM, Joint Laboratory of Optics of Palack\'y
University and Institute of Physics of Academy of Sciences of the
Czech Republic, 17. listopadu 12, 772 07 Olomouc, Czech Republic }

\author{Anton\'{i}n \v{C}ernoch}
\affiliation{RCPTM, Joint Laboratory of Optics of Palack\'y
University and Institute of Physics of Academy of Sciences of the
Czech Republic, 17. listopadu 12, 772 07 Olomouc, Czech Republic }
\affiliation{Institute of Physics of Academy of Science of the
Czech Republic, Joint Laboratory of Optics of Palack\'y University
and Institute of Physics of Academy of Sciences of the Czech
Republic, 17. listopadu 50A, 77207 Olomouc, Czech Republic}

\author{Karel Lemr}
\affiliation{RCPTM, Joint Laboratory of Optics of Palack\'y
University and Institute of Physics of Academy of Sciences of the
Czech Republic, 17. listopadu 12, 772 07 Olomouc, Czech Republic }

\author{Adam Miranowicz}
\affiliation{CEMS, RIKEN, 351-0198 Wako-shi, Japan}
\affiliation{Faculty of Physics, Adam Mickiewicz University,
PL-61-614 Pozna\'n, Poland}

\begin{abstract}
{In standard optical tomographic methods, the off-diagonal
elements of a density matrix $\rho$ are measured indirectly. Thus,
the reconstruction of $\rho$, even if it is based on linear
inversion, typically magnifies small errors in the experimental
data. Recently, an optimal tomography [Phys. Rev. A {\bf 90},
062123 (2014)] has been proposed theoretically to measure
one-by-one all the elements of $\rho$. Thus, the relative errors
in the reconstructed state can be the same as those in the
experimental data. We implemented this method for two-qubit
polarization states performing both local and global measurements.
For comparison, we also experimentally implemented other
well-known tomographic protocols based solely on local
measurements (of, e.g., the Pauli operators and
James-Kwiat-Munro-White projectors) and those with mutually
unbiased bases requiring both local and global measurements. We
reconstructed seventeen of two-qubit polarization states including
separable, partially and maximally entangled. Our experiments show
the highest stability against errors of our method in comparison
to the other quantum tomographies. In particular, we demonstrate
that each optimally-reconstructed state} is embedded in the
uncertainty  circle of the
smallest radius, both in terms of the trace distance and
disturbance. We explain how to estimate experimentally the
uncertainty radii for all the implemented tomographies and show
that, for {each reconstructed state}, the relevant uncertainty
circles intersect indicating the approximate location of
the corresponding physical density matrix.
\end{abstract}



\pacs{03.65.Wj, 03.67.-a, 42.50.Ex}


\maketitle

\textit{Introduction.---} {Quantum tomographic methods are
indispensable tools in experimental quantum physics. Indeed,}
characterizing quantum states and quantum processes are 
essential problems in studying the performance and evolution of
quantum systems~\cite{ParisBook} and in developing quantum
technologies~\cite{Georgescu12}. Moreover, both these problems are mathematically
equivalent and are usually solved by applying quantum state
tomography (QST). This approach is typically based on linear
inversion~\cite{NielsenBook} and maximum-likelihood
estimation~\cite{James01,Rehacek01,Blume10a, Teo11,Teo12,
Smolin12, Halenkova2012, Halenkova2012a, Bart12,Bart13a,Bart14}.
There are also other proposals of quantum state estimation based
on, e.g., least-squares inversion~\cite{Opatrny97} as well as
Bayesian mean estimation~\cite{ParisBook, Blume10b, Huszar12}, or
linear regression estimation~\cite{Qi13}. There {exist dozens
of QST protocols even in the special case of reconstructing the
photonic polarization states (for a review see
Ref.~\cite{Altepeter05review} and also, e.g.,
Refs.~\cite{Burgh08,Bogdanov10a,Adamson10,Sansoni10,Altepeter11,Pryde05,Lundeen11,Lundeen12,Salvail13}).
Thus,} choosing the best of  them appeared to be not a simple
task. However, a recent paper~\cite{Miranowicz14} 
describes an optimal QST protocol that minimizes the condition
number $\kappa$ that characterizes the robustness against
experimental errors. Condition numbers were also used for
investigating the error stability of optical tomographic
protocols in 
Refs.~\cite{Bogdanov10a,Bogdanov10b,Bogdanov11a}. In this Letter
we experimentally study this optimal protocol {in comparison
to four} other popular approaches using the same experimental
setup {for the reconstruction of two polarization-entangled
photons. These protocols include those based solely on the local
measurements of (i) the James-Kwiat-Munro-White (JKMW)
projectors~\cite{James01}, (ii) the Pauli operators, and (iii)
their eigenstates (the so-called standard
basis)~\cite{Altepeter05,Burgh08}, together with (iv) the protocol
of Adamson and Steinberg~\cite{Adamson10} based on
mutually-unbiased bases (MUB) and applying both local and global
measurements, analogously to the optimal protocol.} To compare
these protocols we first derive a relation between the radius of
an error  circle associated with the reconstructed state and
measured quantities. The radii correspond to the trace distance
between the ideal density matrices and the reconstructed noisy
ones. However, they can also be interpreted in terms of fidelity
(or disturbance).

All the approaches analyzed  here are based on solving a
linear-system problem $Ax = b$, where $A$ is referred to as the
\emph{coefficient matrix}, ${b}$ is the \emph{observation vector}
containing the measured data, and $x={\rm vec}(\rho)$ is a real
vector describing the unknown state $\rho$ to be reconstructed.
Here we choose
$$
x={\rm vec}(\rho) = [\rho_{11},{\rm Re} \rho_{12},{\rm Im}
\rho_{12},{\rm Re} \rho_{13},{\rm Im} \rho_{13}, ...,\rho_{44}]^T.
$$
Conversely, a two-qubit density matrix $\rho$ can be represented
as a real vector $x=(x_1,...,x_{16})$ with its elements given as
follows
$$
  \rho(x) = \left[
\begin{array}{cccc}
 x_{1} & x_{2}+i x_{3} & x_{4}+i x_{5} & x_{6}+i x_{7} \\
 x_{2}-i x_{3} & x_{8} & x_{9}+i x_{10} & x_{11}+i x_{12} \\
 x_{4}-i x_{5} & x_{9}-i x_{10} & x_{13} & x_{14}+i x_{15} \\
 x_{6}-i x_{7} & x_{11}-i x_{12} & x_{14}-i x_{15} & x_{16} \\
\end{array}
\right]
$$
The already mentioned condition number $\kappa$ depends only on
$A$, i.e., the system of equations used to estimate the density
matrix from the experimental data $b$. The reliability of the
reconstructed density matrix $\rho$ corresponding to the vector
$x=A^{-1}b$ for a given set of rotations $A$ (representing our
linear tomographic system) and for the measured data $b$ depends
on the value of $\kappa$. To show the operational (or physical)
importance of condition numbers more explicitly, let us recall a
well known theorem (Theorem 8.4 in Ref.~\cite{AtkinsonBook}):
Consider the system $Ax = b$ with nonsingular $A$. Assume
perturbations $\delta b$ in $b$. If perturbations $\delta x$ are
defined implicitly by $A(x+\delta\, x) = b+\delta\, b,$ then it
holds~\cite{AtkinsonBook}:
\begin{equation}\label{eq:Atkinson3}
 \frac{1}{\kappa(A)}
 \frac{||\delta b||}{||b||} \le  \frac{||\delta x||}{||x||} \le \kappa(A)
 \frac{||\delta b||}{||b||}.
\end{equation}
Thus, if the condition number $\kappa(A)$, for a given norm, is
equal (or very close) to one, then small relative changes in the
observation vector $b$ imply equally small relative changes in the
reconstructed state $x$. Here we calculate $
  \kappa(A) \equiv {\rm cond}_2(A)= \sigma_{\max}(A)/{\sigma_{\min}(A)}
$ based on the spectral norm $\Vert A\Vert_{2}$, which is
compatible with Euclidean distance $\|\cdot\|$ used for other
quantities in Eq.~(\ref{eq:Atkinson3}). The norm is defined by the
largest singular value of $A$, i.e, $\Vert A\Vert_{2}=\max[{\rm
svd}(A)]\equiv \sigma_{\max}(A),$ where the function ${\rm
svd}(A)$ returns the singular values of $A$. As shown in
Ref.~\cite{Miranowicz14}, the optimal tomography provides
$\kappa(A)=1 $ for 16 local and nonlocal measurements. By contrast
to this, the  JKMW tomography~\cite{James01} leads to
$\kappa(A)=\sqrt{60.1}$ for 16 local measurements, the standard
separable basis \cite{Altepeter05,Burgh08} yields $\kappa(A)=3$
for 36 local measurements, and mutually-unbiased
bases~\cite{Adamson10,Bandyopadhyay01} gives   $\kappa(A) =
\sqrt{5}$ for 20 local and nonlocal measurements. The tomography
based on Pauli matrices gives $\kappa(A) = \sqrt{2}$ for 16 local
measurements. This suggest that the density matrices reconstructed
with these different protocols reside inside uncertainty 
circles of various radii that depend on $\kappa$.

\vspace{5mm}

\textit{Error analysis.---} From the linearity of the linear
inversion problem and Eq.~(\ref{eq:Atkinson3}) it follows that
\begin{equation}\label{eq:Bnds}
 \frac{1}{\kappa(A)}
 \frac{\|\delta b\|}{||b+\delta b||} \le  \frac{\|\delta x\|}{\|x+\delta x\|} \le \kappa(A)
 \frac{\|\delta b\|}{\|b+\delta b\|}.
\end{equation}
Let us quantify the quality of a tomography protocol with the
trace distance $E\equiv T[\rho(x),\rho(x+\delta
x)]=\tfrac{1}{2}\tr \sqrt{(\delta \rho)^2}$, where
$\delta\rho\equiv\rho(\delta x)$, between the ideal, $\rho(x)$,
and perturbed, $\rho(x+\delta x)$, density matrices. The trace
distance is a proper measure of the distance between two density
matrices and has a statistical interpretation in terms of the
probability of distinguishing between the two matrices. Moreover,
it provides a single number that quantifies the error introduced
by the protocol. We can relate $E$  to $\| \delta x\| $ by using a
standard inequality between the quadratic and arithmetic means of
eigenvalues of $\sqrt{(\delta\rho)^2}$. The result is
\begin{equation}\label{eq:upperB}
2E=\tr \sqrt{(\delta\rho)^2} \le \sqrt{d\,\tr
[(\delta\rho)^{2}]}\le \sqrt{2d}\,\|\delta x\|,
\end{equation}
where, for a two-qubit density matrix, $\tr[(\delta\,\rho)^2] =
2\sum^{16}_{i=1} \delta x^2_i-(\delta x^2_1+\delta x^2_5+\delta
x^2_{13}+\delta x^2_{16})$ and the matrix dimension is $d=4$ . By
combining inequalities in Eqs.~(\ref{eq:Bnds}) and
(\ref{eq:upperB}) we arrive at
\begin{equation}\label{eq:2E}
E \le \sqrt{\frac{d}{2}}\,\kappa(A)\, \frac{\|\delta b\|\|x+\delta
x\|}{\|b+\delta b\|},
\end{equation}
where the random deviations $\delta b$ can be related to the
vector of standard deviations $\sigma(b)$ associated with the mean
values $b$. The distribution of random photon counts $b+\delta b$
is usually described by the Poisson statistics. After performing
the measurements $b+\delta\, b$ one assumes that $b+\delta\,
b\approx b=\sigma^2(b)$, i.e., the measurement outcomes are the
most likely (the mean) number of counts. The relative error of
such approximation is small if the number of counts is large. In
order to compare the robustness of the tomographies, the
deviations $\delta b$ need to be bounded from above. For the
Poisson distribution ,the probability of a magnitude of a random
deviation, $|\delta\,b_i|>2\sqrt{2b_i}$, is given by its
cumulative distribution function ($\mathrm{CDF}$) as
$\mathrm{Pr}(|\delta\,b_i|>2\sqrt{2b_i})=\mathrm{CDF}(x_+)-\mathrm{CDF}(x_-)$,
where $x_{\pm}=\lfloor b_i\pm 2\sqrt{2b_i}\rfloor$ and for the
Poisson distribution $\mathrm{CDF}(x<0)=0$. The probability
$\mathrm{Pr}(|\delta\,b_i|>2\sqrt{2b_i})>0.981$  is very high for
all $b_i$. For $b_i>20$ its value is already
$\mathrm{Pr}(|\delta\,b_i|>2\sqrt{2b_i})>0.993$. The same approach
applied to the Gaussian distribution provides the widely used
$3\sigma$ rule, which tells us that almost certainly (with
probability $0.997$) $|\delta b_i| < 3\sigma(b_i)$. The
statistically justified inequality $|\delta b_i| <
2\sqrt{2}\sigma(b_i)$ leads to
\begin{equation}
\label{eq:R}
 E \le R\equiv 2\sqrt{d}\,\kappa(A)\,
\frac{\|\sigma(b)\|\|x+\delta x\|}{\|b+\delta b\|}.
\end{equation}
We have defined the uncertainty radius of the state estimation
$R$, which is the maximal distance between the state and its
estimate, in terms of only the directly measured quantities. This
is an important result as it allows to directly estimate the
quality of a given state reconstruction in a very convenient way.
As we will demonstrate, it also allows to visually compare the
outcomes of various tomographies. This result allows to easily
characterize the quality of reconstruction without knowing $\rho$
a priori. Moreover, $R$ can be used as a sanity check for the
results of maximum likelihood methods, because the proper density
matrices should be contained within the uncertainty circle of
a radius~$R$.

However, using the uncertainty radius $R$, one  can overestimate
the value of the disturbance $E$. Let us introduce $k=\|\delta b
\|/\|\sigma(b)\|$, where $0 \le k \le 2\sqrt{2}$ for the
Poissonian statistics. In the most general case we can write
\begin{equation}\label{eq:probE}
\frac{kR}{4\sqrt{d}\kappa^2(A)}\le E \le \frac{kR}{2\sqrt{2}},
\end{equation}
where the lower bound is derived with help of  Eqs.~(\ref{eq:2E})
and (\ref{eq:R}), and the relation between the trace distance and
the Hilbert-Schmidt distance
$D_{\mathrm{HS}}(\rho,\rho+\delta\,\rho)=\sqrt{\tr[(\delta\,\rho)^{2}]}\le
2E$. For an arbitrary distribution of the results $b_i$, the
Chebyshev's inequality implies that the probability of finding the
reconstructed state inside of an error  circle of the radius
${kR}/{2\sqrt{2}}$ is bounded from below by $1-1/k^2$. This means
that the minimum of $50\%$ of values must lie within the
$\sqrt{2}$ standard deviations of the mean regardless of the
distribution, i.e., the value of $R/2$ bounds from above the
median of $E$ for any distribution. For the Poisson distribution
we can find a tighter upper bound on the probability of
$b_i+\delta\,b_i>b_i+k\sigma(b_i)$ than the one provided by the
Chebyshev inequality, i.e.,  $\mathrm{Pr}(X>x)\le
\mathrm{e}^{-\mu}(\mathrm{e}\mu/x)^x$ (Theorem 5.4 in
Ref.~\cite{Mitzenmacher}), where $X=b_i+\delta\,b_i$ is the random
variable, $\mu=b_i$ and $x=\mu + k\sqrt{\mu}$.

For characterizing the quality of tomographic protocols we can
also introduce the disturbance $D_B(\rho,\rho+\delta
\rho)=1-[\tr\sqrt{\sqrt{\rho}(\rho+\delta
\rho)\sqrt{\rho}}]^2=1-F(\rho,\rho+\delta\, \rho)$, where $F$ is
the  fidelity related to Bures metrics, which fulfils
$D_B(\rho,\rho+\delta \rho)\le T(\rho,\rho+\delta \rho)$. Thus,
$D_B\le E\le R$. The disturbance was used in Ref.~\cite{Adamson10}
for comparing the results of two-qubit tomographies. However, in
our analysis we apply the trace distance as it provides a more
convenient theoretical framework.

\vspace{5mm}

\begin{figure}
\includegraphics[width=8cm]{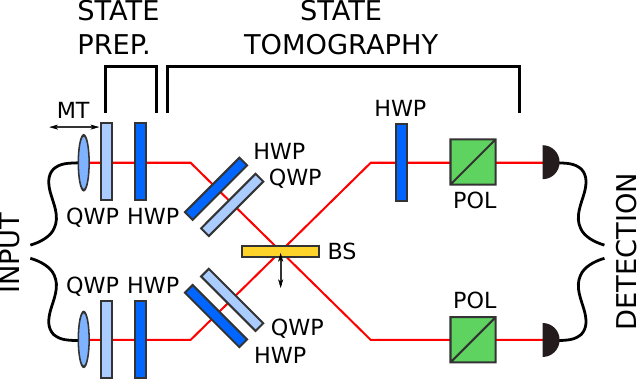}
\caption{\label{fig:setup} Experimental setup performing both
local and nonlocal polarization projections for the four studied
tomographies. Linear-optical components are    quarter-wave plate
(QWP),   half-wave plate (QWP), motorized translation (MT) to
stabilize two-photon overlap,    horizontally retractable balanced
beam splitter (BS), polarizing cube (POL).}
\end{figure}

\textit{Experimental results.---} The results obtained in the
previous section suggest that the error $E$ of state estimation
depends both on $\kappa(A)$ and the measured quantities $b_i$.
Therefore, in order to compare the above-mentioned tomographic
protocols we have prepared 17 two-qubit states and performed four
tomographic protocols on each of them. These states are two-photon
states described in the polarization basis $\lbrace HH,HV,VH,VV
\rbrace$. This  means that, e.g., $\rho_{11}=x_1$ is the
probability of detecting the two photons in the polarization state
$\ket{HH}$ (both photons are polarized horizontally). We have
generated separable and polarization-entangled photon pairs using
the process of spontaneous parametric down-conversion occurring in
a pair of BBO crystals (the so-called Kwiat\etal{}  source
\cite{Kwiat99}). With our source, we have observed about $2.10^3$
two-photon detections per second having 200\,mW of pumping power
at 355\,nm. Generated photons were subsequently brought to the
input of our tomography setup (see Fig.~\ref{fig:setup}), where
the required states were prepared and subsequently both the local
and nonlocal polarizations projections were performed. The
preparation of a given state was achieved using pairs of half
(HWP) and quarter (QWP) wave plates in the input mode of each
photon.

In order to perform local projections on individual photons, we
have shifted the beam splitter BS horizontally so that the
reflections are no longer coupled to the output ports. Then for
each local projection, we have adjusted the HWP and QWP in each
photon's path and then subjected the photons to a polarizing cube.
Respective two-photon detections were registered for 5 seconds.

The nonlocal projections are achieved by combining the local state
transformations using the HWPs and QWPs with the singlet-state
projection on a balanced beam splitter BS. For this procedure to
work, an additional HWP (set to $45^\circ$) has to be placed in
one output mode of the beam splitter before the photons are
subjected to the polarizers. Again, the two-photon detections were
counted for 5 seconds.

While evidently the beam splitter is superfluous for local
projections and the polarizing cubes are  unnecessary for the
nonlocal projections, we maintain all the components in the setup
for all times deliberately since we need to compare the observed
detection rates across local and nonlocal measurements. This would
be problematic without keeping all the components in the setup
since the components introduce different technological losses
(e.g. back-reflections, scattering). Further to that, our setup
allows to switch between the local and nonlocal projections
without much effort.

\begin{figure}
\includegraphics[width=8.5cm]{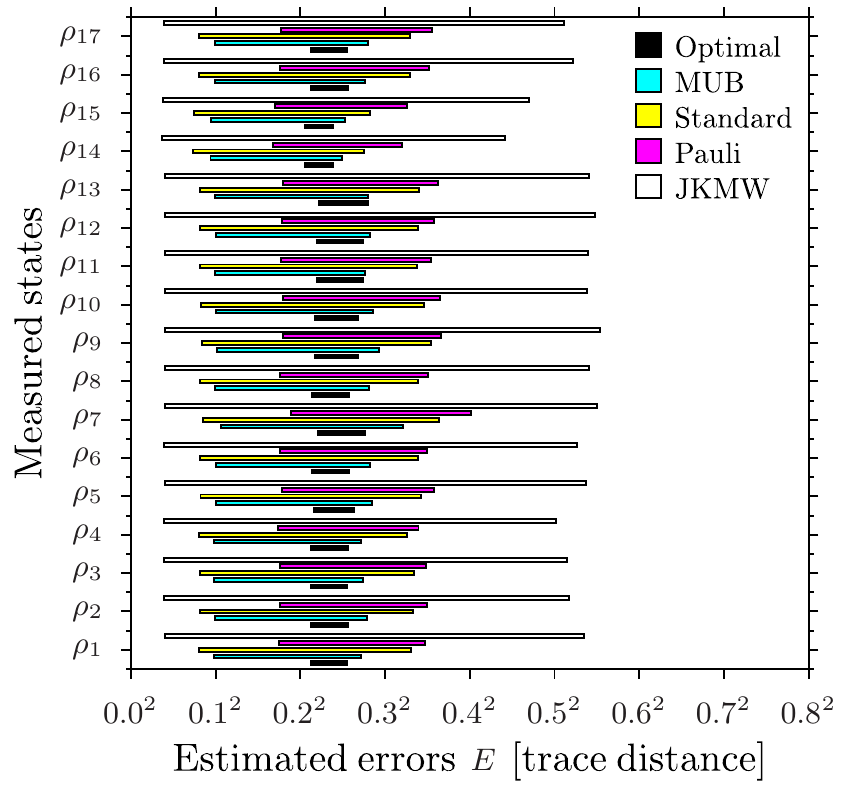}
\caption{\label{fig:E} Experimentally recovered range of possible
errors $E$ for five tomographies {(including the optimal, MUB,
standard-separable-basis, Pauli operators, and JKMW protocols)}
for 17 different two-qubit states. The shaded areas correspond to
the most probable range of the error $E,$ given by
Eq.~(\ref{eq:probE}) for $k=\sqrt{2}$. The maximum error $R$ is
twice the upper limit $r=R/2$ of the plotted error range [see
Eq.~(\ref{eq:R})].}
\end{figure}

For each tomography we have gathered the coincidence counts
$b+\delta b$ for the specific projectors (see the
Supplement~\cite{supplement}). After performing the measurements
we estimated the standard deviations as $b+\delta\, b\approx
b=\sigma^2(b)$. This is justified for large values of $b_i$ as the
relative error of estimating $\sigma(b_i)$ from $b_i+\delta\, b_i$
is $< {(2\sqrt{2}/\sqrt{b_i})^{1/2}}$. In our experiment, we
observe on average that $b_i\approx10^3$ and the smallest values
of $\sigma(b_i)$ do not contribute much to $\|\sigma(b)\|$. Thus,
in order to correct for possible underestimation of
$\|\sigma(b)\|$, we rescale this value by a factor of $1.3$. In
most of the tomographies,  the observation vector $b+\delta b$ is
measured directly. However, for the optimal
tomography~\cite{Miranowicz14} there are 12 measurements which
correspond to the difference of two coincidence counts, say
$c_{i}$ and $c^{\prime}_{i}$. In these cases, the corresponding
entries of the observation vector $b$ are $b_i=\lfloor(c_i-
c^{\prime}_i)/2\rfloor$ are described by the Skellam distribution
(a generalized Poisson distribution), where $\sigma^2(b_i) =
\lfloor(c_i + c^{\prime}_{i})/2\rfloor $. For the Skellam
distribution (similarly to the Poisson distribution) the behaviour
of its cumulative distribution function implies that the largest
disturbance can be limited (with probability $>0.993$ for
$c_i,\,c^{\prime}_i>20$) by $\delta\,b_i\le \lfloor
2\sqrt{2}\sigma (b_i) \rfloor$. The results of our analysis are
shown in Fig.~\ref{fig:E}. The figure demonstrates that the error
range $E$ increases with the condition number $\kappa$ while using
the same setup. This suggests that the optimal tomography can
truly be the best. A more convincing evidence of this observation
is provided by analyzing the relative (trace) distances between
the matrices reconstructed by various tomographies and the sizes
of their error  circles $r=R/2$. We know that the unperturbed
density matrix must be found in the intersection of the error 
circles. This intersection is  very close to the result of the
optimal tomography because it has the smallest error radius. Four
selected representative examples of such geometric construction
are shown in Fig.~\ref{fig:3}.
 Our results for all the reconstructed states can be found in the Supplement~\cite{supplement}.
\begin{figure}
\includegraphics[width=4.25cm]{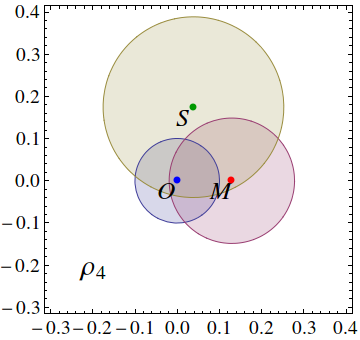}
\includegraphics[width=4.25cm]{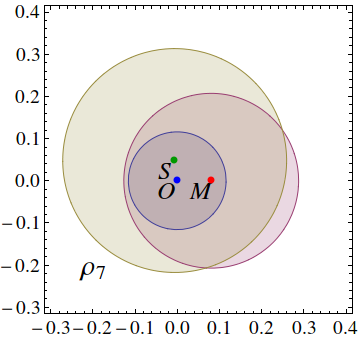}
\includegraphics[width=4.25cm]{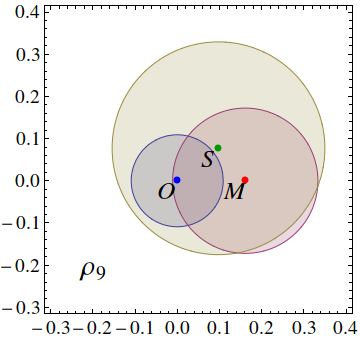}
\includegraphics[width=4.25cm]{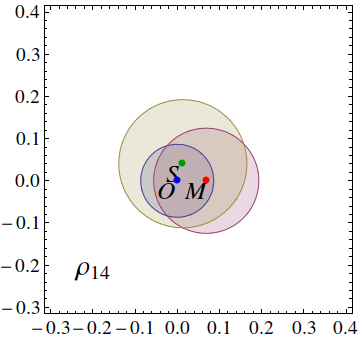}
\caption{\label{fig:3} Relative trace distances between points
corresponding to optimal tomography ($O$),  standard 36 state
tomography ($S$), MUB-based tomography ($M$) representing
reconstructed density matrices and their corresponding disks of
the maximum errors $R$ for four selected reconstructed states.
All the graphically represented distances scaled  in the units of
trace distance. The four reconstructed states  can be approximated
with $\rho_n=\ket{\psi_n}\bra{\psi_n}$, where $\ket{\psi_{4}} =
(\ket{DR}-i\ket{AL})/\sqrt{2}$, $\ket{\psi_{7}} = \ket{HV}$,
$\ket{\psi_{9}} = (\ket{HV}-\ket{VH})/\sqrt{2}$, $\ket{\psi_{14}}
= \ket{e_{1a}e_{1b}}$, where $\ket{e_{1a}} = (-0.6556 +
0.6248i)\ket{H} + 0.4241\ket{V}$ and $\ket{e_{1b}} = (-0.1415 -
0.7165i)\ket{H}+ 0.6831\ket{V}$, where $\ket{H}$, $\ket{V}$,
$\ket{D}$, $\ket{A}$, $\ket{R}$, and $\ket{L}$ stand for the
horizontal, vertical, diagonal, antidiagonal, left-circular, and
right circular states, respectively. An ideally reconstructed
state  lies in the intersection of all the  error disks of radius
$R$.  Note that the discs of radius $R/2,$ associated with the
most probable range of errors $E$ (see Fig.~\ref{fig:E}), do not
necessarily intersect. }
\end{figure}

\vspace{5mm}

\textit{Conclusions.---}  We have for the first time implemented
the optimal two-qubit tomography and compared it with the other
four important tomographic protocols. {This method corresponds
to measuring one by one all the elements $\rho_{nm}$ of a density
matrix $\rho$. This is in contrast to the other protocols, where
the off-diagonal elements of $\rho$ are measured indirectly, i.e.,
the measured photon numbers correspond to linear combinations of
some elements $\rho_{nm}$.} We have developed a method for
estimating the error radii (in units of the trace distance) of
 circles containing the reconstructed density matrices. We
have demonstrated that all the tomographies can be implemented and
compared using the same framework. Our results confirm that the
optimal tomography provides the most reliable results among all
other analyzed protocols. This makes the optimal tomography a
method of choice if the quality of the reconstructed density
matrix is a priority.

\begin{acknowledgments}
\textit{Acknowledgments.---} K.B. acknowledges the support by the
Polish National Science Centre (Grant No. DEC-2013/11/D/ST2/02638)
and by the Foundation for Polish Science (START Programme). K.B.
and A.\v{C}. are supported by the project No. LO1305 of the
Ministry of Education, Youth and Sports of the Czech Republic.
K.L. acknowledges support by the Czech Science Foundation (Grant
No.13-31000P). A.M. is supported by the Polish National Science
Centre under grants DEC-2011/03/B/ST2/01903 and
DEC-2011/02/A/ST2/00305.
\end{acknowledgments}


\clearpage 

\begin{widetext}
\appendix
\begin{center}
{\Large  Priority Choice Experimental Two-qubit Tomography: \\ Measuring
One by One All Elements of Density Matrices:\\ Supplementary Materials}
\end{center}

Here we {show explicitly all the density matrices discussed in
the Letter, which are reconstructed with the optimal tomographic
protocol and those based on: (i) mutually unbiased bases, (ii) the
James-Kwiat-Munro-White projectors, (iii) the tensor products of
the Pauli operators, and (iv) the standard separable basis
corresponding to all the eigenvectors of the Pauli operators.} We
also present the coefficient matrices, observation vectors
corresponding to coincidence counts, the estimated variances for
the observations, and the error radii for each reconstructed
matrix. Finally, we compare the reconstructed matrices
graphically, where we show the relative trace distances between
the reconstructed states and they error radii.

\section{Reconstructed density matrices}

The 17 density matrices are reconstructed by solving linear
inversion problem for four tomographies. We have prepared 17
different states of high purity, which approximately correspond
to:
\begin{equation}\label{eq:states}
\begin{aligned}
\ket{\psi_{1}} = (\ket{HH}-\ket{VV})/\sqrt{2},\qquad&
\ket{\psi_{2}} = (\ket{HH}+\ket{VV})/\sqrt{2},\qquad&
\ket{\psi_{3}} = (\ket{HH}-i\ket{VV})/\sqrt{2},
\\
\ket{\psi_{4}} = (\ket{DR}-i\ket{AL})/\sqrt{2},\qquad&
\ket{\psi_{5}} = (\ket{HV}+i\ket{VH})/\sqrt{2},\qquad&
\ket{\psi_{6}} = (\ket{HV}+\ket{VH})/\sqrt{2},
\\
\ket{\psi_{7}} = \ket{HV},\qquad& \ket{\psi_{8}} =
(\ket{HH}+i\ket{VV})/\sqrt{2},\qquad& \ket{\psi_{9}} =
(\ket{HV}-\ket{VH})/\sqrt{2},
\\
\ket{\psi_{10}} = (\ket{HV}-i\ket{VH})/\sqrt{2},\qquad&
\ket{\psi_{11}} = (\ket{DL}+i\ket{AR})/\sqrt{2},\qquad&
\ket{\psi_{12}} = (\ket{DL}-i\ket{AR})/\sqrt{2},
\\
\ket{\psi_{13}} = \ket{e_{1a}e_{1b}},\qquad& \ket{\psi_{14}} =
\ket{e_{2a}e_{2b}},\qquad& \ket{\psi_{15}} = 0.79\ket{ HV} - 0.61
\ket{VH},
\\
\ket{\psi_{16}} = 0.50\ket{ HV} - 0.87 \ket{VH},\qquad&
\ket{\psi_{17}} = 0.35\ket{ HV} - 0.94 \ket{VH};&
\end{aligned}
\end{equation}
where  $\ket{e_{1a}} = (-0.6556 + 0.6248i)\ket{H} +
0.4241\ket{V}$, $\ket{e_{1b}} = (-0.1415 - 0.7165i)\ket{H}+
0.6831\ket{V}$, $\ket{e_{2a}} = (-0.9608 + 0.2091i)\ket{H} +
0.1822\ket{V}$, and $\ket{e_{2b}} = (0.2613 + 0.7338i)\ket{H}+
0.6271\ket{V}$  are single photon elliptic polarization states. We
mark the data relevant to a particular tomography as follows:
index $O$ for the optimal tomography; $S$ for the standard
36-state tomography; $J$ for the  James-Kwiat-Munro-White
(JKMW) protocol; $M$ for the MUB-based tomography; $P$ for the
Pauli matrices based tomography.

\subsection{Standard 36-state tomography}

$$\rho_{S,1}=
\begin{bmatrix}
  0.4922 &   0.0020 +  0.0156i &  -0.0042 +  0.0354i &  -0.4607 -  0.0750i \\
  0.0020 -  0.0156i &   0.0047 &  -0.0054 +  0.0228i &   0.0255 -  0.0002i \\
 -0.0042 -  0.0354i &  -0.0054 -  0.0228i &   0.0136 &   0.0184 +  0.0656i \\
 -0.4607 +  0.0750i &   0.0255 +  0.0002i &   0.0184 -  0.0656i &   0.4895 \\
\end{bmatrix}
$$

$$\rho_{S,2}=
\begin{bmatrix}
  0.4870 &   0.0029 -  0.0358i &   0.0237 -  0.0103i &   0.4723 +  0.0515i \\
  0.0029 +  0.0358i &   0.0085 &  -0.0000 -  0.0219i &   0.0244 +  0.0413i \\
  0.0237 +  0.0103i &  -0.0000 +  0.0219i &   0.0038 &  -0.0052 -  0.0378i \\
  0.4723 -  0.0515i &   0.0244 -  0.0413i &  -0.0052 +  0.0378i &   0.5007 \\
\end{bmatrix}
$$

$$\rho_{S,3}=
\begin{bmatrix}
  0.5363 &   0.0744 -  0.0611i &   0.0830 -  0.0243i &  -0.0027 +  0.4636i \\
  0.0744 +  0.0611i &   0.0206 &   0.0522 -  0.0036i &  -0.0513 +  0.0602i \\
  0.0830 +  0.0243i &   0.0522 +  0.0036i &   0.0002 &   0.0005 +  0.0402i \\
 -0.0027 -  0.4636i &  -0.0513 -  0.0602i &   0.0005 -  0.0402i &   0.4429 \\
\end{bmatrix}
$$

$$\rho_{S,4}=
\begin{bmatrix}
  0.3005 &   0.2633 +  0.0250i &   0.0456 -  0.2223i &  -0.0648 +  0.2710i \\
  0.2633 -  0.0250i &   0.2482 &   0.0417 -  0.1841i &  -0.0334 +  0.2402i \\
  0.0456 +  0.2223i &   0.0417 +  0.1841i &   0.1270 &  -0.2145 -  0.0463i \\
 -0.0648 -  0.2710i &  -0.0334 -  0.2402i &  -0.2145 +  0.0463i &   0.3244 \\
\end{bmatrix}
$$

$$\rho_{S,5}=
\begin{bmatrix}
  0.0135 &  -0.0859 +  0.0013i &   0.0334 +  0.0685i &   0.0159 +  0.0018i \\
 -0.0859 -  0.0013i &   0.5111 &   0.0397 -  0.4647i &  -0.0002 -  0.0441i \\
  0.0334 -  0.0685i &   0.0397 +  0.4647i &   0.4697 &   0.0078 -  0.0236i \\
  0.0159 -  0.0018i &  -0.0002 +  0.0441i &   0.0078 +  0.0236i &   0.0056 \\
\end{bmatrix}
$$

$$\rho_{S,6}=
\begin{bmatrix}
  0.0146 &   0.0450 +  0.0866i &   0.0849 +  0.0819i &  -0.0046 +  0.0341i \\
  0.0450 -  0.0866i &   0.4606 &   0.4608 -  0.0160i &  -0.0222 -  0.0282i \\
  0.0849 -  0.0819i &   0.4608 +  0.0160i &   0.5177 &  -0.0526 -  0.0125i \\
 -0.0046 -  0.0341i &  -0.0222 +  0.0282i &  -0.0526 +  0.0125i &   0.0072 \\
\end{bmatrix}
$$

$$\rho_{S,7}=
\begin{bmatrix}
  0.0005 &   0.0578 -  0.0482i &   0.0064 +  0.0028i &  -0.0088 -  0.0005i \\
  0.0578 +  0.0482i &   0.9915 &   0.0290 +  0.0678i &  -0.0436 -  0.1064i \\
  0.0064 -  0.0028i &   0.0290 -  0.0678i &   0.0049 &  -0.0043 -  0.0022i \\
 -0.0088 +  0.0005i &  -0.0436 +  0.1064i &  -0.0043 +  0.0022i &   0.0032 \\
\end{bmatrix}
$$

$$\rho_{S,8}=
\begin{bmatrix}
  0.5609 &  -0.0543 +  0.0315i &   0.0357 -  0.0364i &   0.0027 -  0.4704i \\
 -0.0543 -  0.0315i &   0.0067 &  -0.0470 +  0.0079i &  -0.0016 +  0.0511i \\
  0.0357 +  0.0364i &  -0.0470 -  0.0079i &  -0.0091 &   0.0065 -  0.0126i \\
  0.0027 +  0.4704i &  -0.0016 -  0.0511i &   0.0065 +  0.0126i &   0.4416 \\
\end{bmatrix}
$$

$$\rho_{S,9}=
\begin{bmatrix}
 -0.0208 &  -0.0024 -  0.0670i &  -0.0036 +  0.0319i &  -0.0224 -  0.0403i \\
 -0.0024 +  0.0670i &   0.5767 &  -0.4584 -  0.0718i &  -0.0076 -  0.0478i \\
 -0.0036 -  0.0319i &  -0.4584 +  0.0718i &   0.4334 &   0.0134 +  0.0045i \\
 -0.0224 +  0.0403i &  -0.0076 +  0.0478i &   0.0134 -  0.0045i &   0.0107 \\
\end{bmatrix}
$$

$$\rho_{S,10}=
\begin{bmatrix}
 -0.0118 &   0.0243 -  0.0103i &   0.0134 +  0.0063i &  -0.0090 -  0.0064i \\
  0.0243 +  0.0103i &   0.5080 &   0.0499 +  0.4684i &  -0.0174 -  0.0050i \\
  0.0134 -  0.0063i &   0.0499 -  0.4684i &   0.4801 &   0.0302 +  0.0537i \\
 -0.0090 +  0.0064i &  -0.0174 +  0.0050i &   0.0302 -  0.0537i &   0.0237 \\
\end{bmatrix}
$$

$$\rho_{S,11}=
\begin{bmatrix}
  0.2826 &   0.2502 +  0.0153i &  -0.0157 +  0.2358i &   0.0053 -  0.2593i \\
  0.2502 -  0.0153i &   0.2221 &  -0.0295 +  0.2241i &   0.0222 -  0.2349i \\
 -0.0157 -  0.2358i &  -0.0295 -  0.2241i &   0.2718 &  -0.2432 +  0.0003i \\
  0.0053 +  0.2593i &   0.0222 +  0.2349i &  -0.2432 -  0.0003i &   0.2235 \\
\end{bmatrix}
$$

$$\rho_{S,12}=
\begin{bmatrix}
  0.2100 &  -0.2611 -  0.0385i &  -0.0419 -  0.2516i &   0.0345 -  0.2156i \\
 -0.2611 +  0.0385i &   0.2815 &   0.0452 +  0.2460i &   0.0115 +  0.2405i \\
 -0.0419 +  0.2516i &   0.0452 -  0.2460i &   0.2312 &   0.2259 +  0.0926i \\
  0.0345 +  0.2156i &   0.0115 -  0.2405i &   0.2259 -  0.0926i &   0.2772 \\
\end{bmatrix}
$$

$$\rho_{S,13}=
\begin{bmatrix}
  0.2849 &  -0.2477 -  0.0400i &   0.0073 +  0.2296i &   0.0167 +  0.2654i \\
 -0.2477 +  0.0400i &   0.2042 &   0.0104 -  0.2183i &  -0.0427 -  0.2274i \\
  0.0073 -  0.2296i &   0.0104 +  0.2183i &   0.2687 &   0.2297 -  0.0240i \\
  0.0167 -  0.2654i &  -0.0427 +  0.2274i &   0.2297 +  0.0240i &   0.2422 \\
\end{bmatrix}
$$

$$\rho_{S,14}=
\begin{bmatrix}
  0.3786 &  -0.0779 -  0.3764i &  -0.1289 +  0.1049i &   0.1243 +  0.0775i \\
 -0.0779 +  0.3764i &   0.3745 &  -0.0776 -  0.1422i &  -0.0950 +  0.1000i \\
 -0.1289 -  0.1049i &  -0.0776 +  0.1422i &   0.1345 &  -0.0377 -  0.1163i \\
  0.1243 -  0.0775i &  -0.0950 -  0.1000i &  -0.0377 +  0.1163i &   0.1124 \\
\end{bmatrix}
$$

$$\rho_{S,15}=
\begin{bmatrix}
  0.5241 &   0.1701 +  0.4098i &  -0.0732 +  0.0497i &  -0.0613 -  0.0740i \\
  0.1701 -  0.4098i &   0.3733 &   0.0048 +  0.0800i &  -0.0776 +  0.0176i \\
 -0.0732 -  0.0497i &   0.0048 -  0.0800i &   0.0586 &   0.0165 +  0.0519i \\
 -0.0613 +  0.0740i &  -0.0776 -  0.0176i &   0.0165 -  0.0519i &   0.0440 \\
\end{bmatrix}
$$

$$\rho_{S,16}=
\begin{bmatrix}
  0.0079 &  -0.0088 -  0.1191i &  -0.0026 +  0.0469i &  -0.0179 -  0.0369i \\
 -0.0088 +  0.1191i &   0.6443 &  -0.4307 -  0.0597i &  -0.0144 -  0.0826i \\
 -0.0026 -  0.0469i &  -0.4307 +  0.0597i &   0.3254 &   0.0248 +  0.0298i \\
 -0.0179 +  0.0369i &  -0.0144 +  0.0826i &   0.0248 -  0.0298i &   0.0224 \\
\end{bmatrix}
$$

$$\rho_{S,17}=
\begin{bmatrix}
  0.0071 &  -0.0098 -  0.1224i &  -0.0079 +  0.0450i &  -0.0180 -  0.0331i \\
 -0.0098 +  0.1224i &   0.7482 &  -0.3769 -  0.0542i &  -0.0215 -  0.1008i \\
 -0.0079 -  0.0450i &  -0.3769 +  0.0542i &   0.2232 &   0.0274 +  0.0197i \\
 -0.0180 +  0.0331i &  -0.0215 +  0.1008i &   0.0274 -  0.0197i &   0.0215 \\
\end{bmatrix}
$$

\subsection{ JKMW Tomography}

$$\rho_{J,1}=
\begin{bmatrix}
  0.4879 &  -0.0241 +  0.0194i &  -0.0198 +  0.0473i &  -0.4503 -  0.0438i \\
 -0.0241 -  0.0194i &   0.0054 &  -0.0313 +  0.1193i &   0.0428 -  0.0066i \\
 -0.0198 -  0.0473i &  -0.0313 -  0.1193i &   0.0225 &  -0.0023 +  0.0852i \\
 -0.4503 +  0.0438i &   0.0428 +  0.0066i &  -0.0023 -  0.0852i &   0.4842 \\
\end{bmatrix}
$$

$$\rho_{J,2}=
\begin{bmatrix}
  0.4748 &   0.0156 -  0.0702i &  -0.0009 +  0.0089i &   0.4543 +  0.0190i \\
  0.0156 +  0.0702i &   0.0107 &   0.0457 -  0.1052i &   0.0248 +  0.0104i \\
 -0.0009 -  0.0089i &   0.0457 +  0.1052i &   0.0079 &   0.0098 -  0.0398i \\
  0.4543 -  0.0190i &   0.0248 -  0.0104i &   0.0098 +  0.0398i &   0.5065 \\
\end{bmatrix}
$$

$$\rho_{J,3}=
\begin{bmatrix}
  0.5221 &   0.0753 -  0.0707i &   0.0706 -  0.0238i &   0.0314 +  0.4451i \\
  0.0753 +  0.0707i &   0.0213 &   0.0016 -  0.0338i &  -0.0532 +  0.0446i \\
  0.0706 +  0.0238i &   0.0016 +  0.0338i &   0.0112 &   0.0008 +  0.0618i \\
  0.0314 -  0.4451i &  -0.0532 -  0.0446i &   0.0008 -  0.0618i &   0.4454 \\
\end{bmatrix}
$$

$$\rho_{J,4}=
\begin{bmatrix}
  0.2807 &   0.2535 +  0.0302i &   0.0392 -  0.1853i &  -0.0417 +  0.2515i \\
  0.2535 -  0.0302i &   0.2543 &  -0.0142 -  0.1473i &  -0.0489 +  0.1999i \\
  0.0392 +  0.1853i &  -0.0142 +  0.1473i &   0.1256 &  -0.2106 -  0.0075i \\
 -0.0417 -  0.2515i &  -0.0489 -  0.1999i &  -0.2106 +  0.0075i &   0.3394 \\
\end{bmatrix}
$$

$$\rho_{J,5}=
\begin{bmatrix}
  0.0260 &  -0.0955 +  0.0352i &   0.0370 +  0.0450i &   0.0842 +  0.0143i \\
 -0.0955 -  0.0352i &   0.5035 &   0.0139 -  0.4182i &  -0.0177 -  0.0134i \\
  0.0370 -  0.0450i &   0.0139 +  0.4182i &   0.4676 &  -0.0199 -  0.0215i \\
  0.0842 -  0.0143i &  -0.0177 +  0.0134i &  -0.0199 +  0.0215i &   0.0029 \\
\end{bmatrix}
$$

$$\rho_{J,6}=
\begin{bmatrix}
  0.0367 &   0.0389 +  0.1059i &   0.0675 +  0.0395i &   0.0312 +  0.0146i \\
  0.0389 -  0.1059i &   0.4500 &   0.4370 -  0.0716i &  -0.0533 -  0.0099i \\
  0.0675 -  0.0395i &   0.4370 +  0.0716i &   0.5064 &  -0.0482 -  0.0141i \\
  0.0312 -  0.0146i &  -0.0533 +  0.0099i &  -0.0482 +  0.0141i &   0.0069 \\
\end{bmatrix}
$$

$$\rho_{J,7}=
\begin{bmatrix}
  0.0062 &   0.0456 -  0.0314i &   0.0057 +  0.0030i &   0.0072 -  0.0079i \\
  0.0456 +  0.0314i &   0.9818 &   0.0226 +  0.0910i &  -0.0969 -  0.0714i \\
  0.0057 -  0.0030i &   0.0226 -  0.0910i &   0.0032 &  -0.0035 +  0.0032i \\
  0.0072 +  0.0079i &  -0.0969 +  0.0714i &  -0.0035 -  0.0032i &   0.0087 \\
\end{bmatrix}
$$

$$\rho_{J,8}=
\begin{bmatrix}
  0.5550 &  -0.0784 +  0.0200i &  -0.0058 -  0.0372i &  -0.0223 -  0.4409i \\
 -0.0784 -  0.0200i &   0.0135 &   0.0219 +  0.0377i &  -0.0032 +  0.0374i \\
 -0.0058 +  0.0372i &   0.0219 -  0.0377i &   0.0043 &  -0.0027 -  0.0162i \\
 -0.0223 +  0.4409i &  -0.0032 -  0.0374i &  -0.0027 +  0.0162i &   0.4271 \\
\end{bmatrix}
$$

$$\rho_{J,9}=
\begin{bmatrix}
  0.0089 &  -0.0324 -  0.0301i &  -0.0031 -  0.0189i &   0.0009 -  0.0043i \\
 -0.0324 +  0.0301i &   0.5684 &  -0.4348 +  0.0409i &  -0.0166 -  0.0446i \\
 -0.0031 +  0.0189i &  -0.4348 -  0.0409i &   0.4209 &   0.0002 -  0.0151i \\
  0.0009 +  0.0043i &  -0.0166 +  0.0446i &   0.0002 +  0.0151i &   0.0018 \\
\end{bmatrix}
$$

$$\rho_{J,10}=
\begin{bmatrix}
  0.0058 &   0.0400 -  0.0019i &   0.0128 -  0.0240i &  -0.0401 -  0.0454i \\
  0.0400 +  0.0019i &   0.5133 &   0.0953 +  0.3781i &  -0.0293 +  0.0144i \\
  0.0128 +  0.0240i &   0.0953 -  0.3781i &   0.4753 &   0.0619 +  0.0094i \\
 -0.0401 +  0.0454i &  -0.0293 -  0.0144i &   0.0619 -  0.0094i &   0.0056 \\
\end{bmatrix}
$$

$$\rho_{J,11}=
\begin{bmatrix}
  0.2918 &   0.2453 -  0.0200i &  -0.0285 +  0.1767i &  -0.0527 -  0.2169i \\
  0.2453 +  0.0200i &   0.1992 &   0.0202 +  0.2358i &   0.0220 -  0.2014i \\
 -0.0285 -  0.1767i &   0.0202 -  0.2358i &   0.2948 &  -0.2464 +  0.0008i \\
 -0.0527 +  0.2169i &   0.0220 +  0.2014i &  -0.2464 -  0.0008i &   0.2142 \\
\end{bmatrix}
$$

$$\rho_{J,12}=
\begin{bmatrix}
  0.1923 &  -0.2367 -  0.0390i &  -0.0338 -  0.1873i &  -0.0453 -  0.2798i \\
 -0.2367 +  0.0390i &   0.3002 &   0.0996 +  0.1912i &   0.0039 +  0.2219i \\
 -0.0338 +  0.1873i &   0.0996 -  0.1912i &   0.2127 &   0.2404 +  0.0740i \\
 -0.0453 +  0.2798i &   0.0039 -  0.2219i &   0.2404 -  0.0740i &   0.2947 \\
\end{bmatrix}
$$

$$\rho_{J,13}=
\begin{bmatrix}
  0.3064 &  -0.2350 -  0.0157i &   0.0162 +  0.1740i &   0.0769 +  0.2162i \\
 -0.2350 +  0.0157i &   0.1884 &  -0.0753 -  0.2582i &  -0.0327 -  0.1833i \\
  0.0162 -  0.1740i &  -0.0753 +  0.2582i &   0.2840 &   0.2538 -  0.0131i \\
  0.0769 -  0.2162i &  -0.0327 +  0.1833i &   0.2538 +  0.0131i &   0.2211 \\
\end{bmatrix}
$$

$$\rho_{J,14}=
\begin{bmatrix}
  0.3803 &  -0.0676 -  0.4027i &  -0.1333 +  0.1142i &   0.1197 +  0.1053i \\
 -0.0676 +  0.4027i &   0.3748 &  -0.0745 -  0.1464i &  -0.0932 +  0.1079i \\
 -0.1333 -  0.1142i &  -0.0745 +  0.1464i &   0.1345 &  -0.0335 -  0.1211i \\
  0.1197 -  0.1053i &  -0.0932 -  0.1079i &  -0.0335 +  0.1211i &   0.1104 \\
\end{bmatrix}
$$

$$\rho_{J,15}=
\begin{bmatrix}
  0.5276 &   0.1690 +  0.4181i &  -0.0627 +  0.0920i &  -0.0903 -  0.0653i \\
  0.1690 -  0.4181i &   0.3819 &   0.0342 +  0.0807i &  -0.0764 +  0.0379i \\
 -0.0627 -  0.0920i &   0.0342 -  0.0807i &   0.0478 &   0.0032 +  0.0440i \\
 -0.0903 +  0.0653i &  -0.0764 -  0.0379i &   0.0032 -  0.0440i &   0.0427 \\
\end{bmatrix}
$$

$$\rho_{J,16}=
\begin{bmatrix}
  0.0207 &  -0.0045 -  0.1146i &  -0.0112 +  0.0073i &  -0.0325 -  0.0020i \\
 -0.0045 +  0.1146i &   0.6312 &  -0.4297 +  0.0089i &   0.0034 -  0.0464i \\
 -0.0112 -  0.0073i &  -0.4297 -  0.0089i &   0.3450 &   0.0040 +  0.0183i \\
 -0.0325 +  0.0020i &   0.0034 +  0.0464i &   0.0040 -  0.0183i &   0.0031 \\
\end{bmatrix}
$$

$$\rho_{J,17}=
\begin{bmatrix}
  0.0202 &  -0.0172 -  0.1220i &  -0.0095 +  0.0268i &  -0.0629 -  0.0175i \\
 -0.0172 +  0.1220i &   0.7415 &  -0.3322 -  0.0029i &  -0.0138 -  0.0503i \\
 -0.0095 -  0.0268i &  -0.3322 +  0.0029i &   0.2341 &   0.0108 -  0.0009i \\
 -0.0629 +  0.0175i &  -0.0138 +  0.0503i &   0.0108 +  0.0009i &   0.0042 \\
\end{bmatrix}
$$

\subsection{MUB-based tomography}

$$\rho_{M,1}=
\begin{bmatrix}
  0.4789 &   0.0857 +  0.0275i &   0.0044 +  0.0534i &  -0.4699 -  0.0476i \\
  0.0857 -  0.0275i &   0.0311 &  -0.0110 -  0.0029i &   0.0332 -  0.0196i \\
  0.0044 -  0.0534i &  -0.0110 +  0.0029i &  -0.0015 &   0.0215 +  0.0758i \\
 -0.4699 +  0.0476i &   0.0332 +  0.0196i &   0.0215 -  0.0758i &   0.4915 \\
\end{bmatrix}
$$

$$\rho_{M,2}=
\begin{bmatrix}
  0.4888 &  -0.0083 -  0.0456i &   0.0198 +  0.0013i &   0.4919 +  0.0541i \\
 -0.0083 +  0.0456i &   0.0296 &   0.0027 -  0.0241i &   0.0206 +  0.0291i \\
  0.0198 -  0.0013i &   0.0027 +  0.0241i &  -0.0322 &  -0.0820 -  0.0477i \\
  0.4919 -  0.0541i &   0.0206 -  0.0291i &  -0.0820 +  0.0477i &   0.5138 \\
\end{bmatrix}
$$

$$\rho_{M,3}=
\begin{bmatrix}
  0.5214 &   0.0980 -  0.0452i &   0.0865 -  0.0535i &   0.0541 +  0.4566i \\
  0.0980 +  0.0452i &   0.0573 &  -0.0021 +  0.0016i &  -0.0472 +  0.0870i \\
  0.0865 +  0.0535i &  -0.0021 -  0.0016i &   0.0053 &  -0.0563 +  0.0557i \\
  0.0541 -  0.4566i &  -0.0472 -  0.0870i &  -0.0563 -  0.0557i &   0.4160 \\
\end{bmatrix}
$$

$$\rho_{M,4}=
\begin{bmatrix}
  0.2423 &   0.2560 +  0.0449i &   0.0503 -  0.2014i &   0.0328 +  0.2533i \\
  0.2560 -  0.0449i &   0.3304 &   0.1157 -  0.1609i &  -0.0336 +  0.2319i \\
  0.0503 +  0.2014i &   0.1157 +  0.1609i &   0.1676 &  -0.1754 -  0.0308i \\
  0.0328 -  0.2533i &  -0.0336 -  0.2319i &  -0.1754 +  0.0308i &   0.2597 \\
\end{bmatrix}
$$

$$\rho_{M,5}=
\begin{bmatrix}
 -0.0004 &  -0.0941 +  0.0188i &   0.0384 +  0.0899i &   0.0017 -  0.0313i \\
 -0.0941 -  0.0188i &   0.4975 &   0.1830 -  0.4264i &   0.0052 -  0.0638i \\
  0.0384 -  0.0899i &   0.1830 +  0.4264i &   0.5160 &   0.0302 -  0.0058i \\
  0.0017 +  0.0313i &   0.0052 +  0.0638i &   0.0302 +  0.0058i &  -0.0131 \\
\end{bmatrix}
$$

$$\rho_{M,6}=
\begin{bmatrix}
  0.0119 &   0.0116 +  0.0964i &   0.0767 +  0.0744i &   0.0106 +  0.0150i \\
  0.0116 -  0.0964i &   0.4477 &   0.4659 +  0.0031i &  -0.0301 -  0.0330i \\
  0.0767 -  0.0744i &   0.4659 -  0.0031i &   0.5525 &  -0.0780 -  0.0024i \\
  0.0106 -  0.0150i &  -0.0301 +  0.0330i &  -0.0780 +  0.0024i &  -0.0121 \\
\end{bmatrix}
$$

$$\rho_{M,7}=
\begin{bmatrix}
  0.0377 &   0.0294 -  0.0443i &   0.0050 -  0.0081i &  -0.0075 +  0.0315i \\
  0.0294 +  0.0443i &   0.9628 &   0.0006 +  0.0373i &  -0.0462 -  0.1064i \\
  0.0050 +  0.0081i &   0.0006 -  0.0373i &   0.0047 &   0.0296 +  0.0029i \\
 -0.0075 -  0.0315i &  -0.0462 +  0.1064i &   0.0296 -  0.0029i &  -0.0052 \\
\end{bmatrix}
$$

$$\rho_{M,8}=
\begin{bmatrix}
  0.5549 &  -0.0058 +  0.0214i &   0.0324 +  0.0106i &  -0.0598 -  0.4494i \\
 -0.0058 -  0.0214i &   0.0236 &   0.0055 -  0.0171i &  -0.0052 +  0.0074i \\
  0.0324 -  0.0106i &   0.0055 +  0.0171i &  -0.0329 &  -0.0199 -  0.0231i \\
 -0.0598 +  0.4494i &  -0.0052 -  0.0074i &  -0.0199 +  0.0231i &   0.4544 \\
\end{bmatrix}
$$

$$\rho_{M,9}=
\begin{bmatrix}
 -0.0157 &  -0.0020 -  0.0764i &  -0.0028 +  0.0180i &  -0.0136 -  0.0240i \\
 -0.0020 +  0.0764i &   0.5405 &  -0.4696 -  0.0885i &  -0.0067 -  0.0289i \\
 -0.0028 -  0.0180i &  -0.4696 +  0.0885i &   0.4525 &   0.0798 -  0.0047i \\
 -0.0136 +  0.0240i &  -0.0067 +  0.0289i &   0.0798 +  0.0047i &   0.0228 \\
\end{bmatrix}
$$

$$\rho_{M,10}=
\begin{bmatrix}
 -0.0061 &  -0.0415 -  0.0233i &   0.0029 -  0.0435i &  -0.0009 -  0.0034i \\
 -0.0415 +  0.0233i &   0.4724 &  -0.0159 +  0.4682i &  -0.0280 +  0.0304i \\
  0.0029 +  0.0435i &  -0.0159 -  0.4682i &   0.4984 &   0.0302 +  0.0410i \\
 -0.0009 +  0.0034i &  -0.0280 -  0.0304i &   0.0302 -  0.0410i &   0.0353 \\
\end{bmatrix}
$$

$$\rho_{M,11}=
\begin{bmatrix}
  0.3090 &   0.2536 -  0.0020i &  -0.0236 +  0.2387i &   0.0049 -  0.2545i \\
  0.2536 +  0.0020i &   0.2169 &  -0.0266 +  0.2206i &   0.0128 -  0.2344i \\
 -0.0236 -  0.2387i &  -0.0266 -  0.2206i &   0.2625 &  -0.2126 -  0.0164i \\
  0.0049 +  0.2545i &   0.0128 +  0.2344i &  -0.2126 +  0.0164i &   0.2116 \\
\end{bmatrix}
$$

$$\rho_{M,12}=
\begin{bmatrix}
  0.2210 &  -0.2802 -  0.0510i &  -0.0475 -  0.2454i &  -0.0166 -  0.2218i \\
 -0.2802 +  0.0510i &   0.2512 &   0.0112 +  0.2530i &   0.0076 +  0.2359i \\
 -0.0475 +  0.2454i &   0.0112 -  0.2530i &   0.1993 &   0.1977 +  0.0841i \\
 -0.0166 +  0.2218i &   0.0076 -  0.2359i &   0.1977 -  0.0841i &   0.3285 \\
\end{bmatrix}
$$

$$\rho_{M,13}=
\begin{bmatrix}
  0.3059 &  -0.2612 -  0.0147i &   0.0146 +  0.2246i &   0.0422 +  0.2544i \\
 -0.2612 +  0.0147i &   0.1762 &  -0.0175 -  0.2085i &  -0.0342 -  0.2370i \\
  0.0146 -  0.2246i &  -0.0175 +  0.2085i &   0.2792 &   0.1911 +  0.0008i \\
  0.0422 -  0.2544i &  -0.0342 +  0.2370i &   0.1911 -  0.0008i &   0.2387 \\
\end{bmatrix}
$$

$$\rho_{M,14}=
\begin{bmatrix}
  0.4095 &  -0.0753 -  0.3984i &  -0.1351 +  0.1251i &   0.1159 +  0.0932i \\
 -0.0753 +  0.3984i &   0.3649 &  -0.0998 -  0.1619i &  -0.0991 +  0.1004i \\
 -0.1351 -  0.1251i &  -0.0998 +  0.1619i &   0.1005 &  -0.0384 -  0.1226i \\
  0.1159 -  0.0932i &  -0.0991 -  0.1004i &  -0.0384 +  0.1226i &   0.1251 \\
\end{bmatrix}
$$

$$\rho_{M,15}=
\begin{bmatrix}
  0.5590 &   0.1688 +  0.4115i &  -0.0750 +  0.0461i &  -0.0342 -  0.0741i \\
  0.1688 -  0.4115i &   0.3472 &   0.0472 +  0.0803i &  -0.0795 +  0.0198i \\
 -0.0750 -  0.0461i &   0.0472 -  0.0803i &   0.0258 &   0.0397 +  0.0409i \\
 -0.0342 +  0.0741i &  -0.0795 -  0.0198i &   0.0397 -  0.0409i &   0.0680 \\
\end{bmatrix}
$$

$$\rho_{M,16}=
\begin{bmatrix}
  0.0095 &  -0.0057 -  0.1331i &  -0.0030 +  0.0330i &  -0.0239 -  0.0074i \\
 -0.0057 +  0.1331i &   0.6278 &  -0.3716 -  0.0931i &  -0.0152 -  0.0669i \\
 -0.0030 -  0.0330i &  -0.3716 +  0.0931i &   0.3255 &   0.0919 +  0.0219i \\
 -0.0239 +  0.0074i &  -0.0152 +  0.0669i &   0.0919 -  0.0219i &   0.0372 \\
\end{bmatrix}
$$

$$\rho_{M,17}=
\begin{bmatrix}
  0.0219 &  -0.0122 -  0.1428i &  -0.0034 +  0.0132i &  -0.0230 -  0.0034i \\
 -0.0122 +  0.1428i &   0.7250 &  -0.3510 -  0.0873i &  -0.0176 -  0.0751i \\
 -0.0034 -  0.0132i &  -0.3510 +  0.0873i &   0.2151 &   0.0860 +  0.0050i \\
 -0.0230 +  0.0034i &  -0.0176 +  0.0751i &   0.0860 -  0.0050i &   0.0379 \\
\end{bmatrix}
$$

\subsection{Optimal tomography}

$$\rho_{O,1}=
\begin{bmatrix}
  0.4879 &  -0.0194 +  0.0278i &   0.0045 +  0.0413i &  -0.4760 -  0.0086i \\
 -0.0194 -  0.0278i &   0.0054 &  -0.0112 -  0.0003i &   0.0336 +  0.0064i \\
  0.0045 -  0.0413i &  -0.0112 +  0.0003i &   0.0225 &  -0.0033 +  0.0768i \\
 -0.4760 +  0.0086i &   0.0336 -  0.0064i &  -0.0033 -  0.0768i &   0.4842 \\
\end{bmatrix}
$$

$$\rho_{O,2}=
\begin{bmatrix}
  0.4748 &   0.0191 -  0.0443i &   0.0193 -  0.0130i &   0.4781 +  0.0019i \\
  0.0191 +  0.0443i &   0.0107 &   0.0026 +  0.0001i &   0.0200 +  0.0379i \\
  0.0193 +  0.0130i &   0.0026 -  0.0001i &   0.0079 &   0.0111 -  0.0464i \\
  0.4781 -  0.0019i &   0.0200 -  0.0379i &   0.0111 +  0.0464i &   0.5065 \\
\end{bmatrix}
$$

$$\rho_{O,3}=
\begin{bmatrix}
  0.5221 &   0.0816 -  0.0442i &   0.0847 -  0.0240i &   0.0530 +  0.4532i \\
  0.0816 +  0.0442i &   0.0213 &  -0.0020 -  0.0026i &  -0.0463 +  0.0584i \\
  0.0847 +  0.0240i &  -0.0020 +  0.0026i &   0.0112 &   0.0095 +  0.0545i \\
  0.0530 -  0.4532i &  -0.0463 -  0.0584i &   0.0095 -  0.0545i &   0.4454 \\
\end{bmatrix}
$$

$$\rho_{O,4}=
\begin{bmatrix}
  0.2807 &   0.2569 +  0.0406i &   0.0456 -  0.2118i &   0.0297 +  0.2155i \\
  0.2569 -  0.0406i &   0.2543 &   0.1047 -  0.1856i &  -0.0304 +  0.2331i \\
  0.0456 +  0.2118i &   0.1047 +  0.1856i &   0.1256 &  -0.2027 -  0.0279i \\
  0.0297 -  0.2155i &  -0.0304 -  0.2331i &  -0.2027 +  0.0279i &   0.3394 \\
\end{bmatrix}
$$

$$\rho_{O,5}=
\begin{bmatrix}
  0.0260 &  -0.0969 +  0.0186i &   0.0380 +  0.0730i &   0.0017 -  0.0078i \\
 -0.0969 -  0.0186i &   0.5035 &   0.1813 -  0.4222i &   0.0051 -  0.0374i \\
  0.0380 -  0.0730i &   0.1813 +  0.4222i &   0.4676 &  -0.0051 -  0.0057i \\
  0.0017 +  0.0078i &   0.0051 +  0.0374i &  -0.0051 +  0.0057i &   0.0029 \\
\end{bmatrix}
$$

$$\rho_{O,6}=
\begin{bmatrix}
  0.0367 &   0.0502 +  0.0923i &   0.0734 +  0.0820i &   0.0102 -  0.0178i \\
  0.0502 -  0.0923i &   0.4500 &   0.4460 -  0.0748i &  -0.0288 -  0.0230i \\
  0.0734 -  0.0820i &   0.4460 +  0.0748i &   0.5064 &  -0.0430 -  0.0023i \\
  0.0102 +  0.0178i &  -0.0288 +  0.0230i &  -0.0430 +  0.0023i &   0.0069 \\
\end{bmatrix}
$$

$$\rho_{O,7}=
\begin{bmatrix}
  0.0062 &   0.0557 -  0.0422i &   0.0047 +  0.0036i &  -0.0072 +  0.0029i \\
  0.0557 +  0.0422i &   0.9818 &   0.0006 +  0.0380i &  -0.0441 -  0.1031i \\
  0.0047 -  0.0036i &   0.0006 -  0.0380i &   0.0032 &  -0.0049 +  0.0027i \\
 -0.0072 -  0.0029i &  -0.0441 +  0.1031i &  -0.0049 -  0.0027i &   0.0087 \\
\end{bmatrix}
$$

$$\rho_{O,8}=
\begin{bmatrix}
  0.5550 &  -0.0622 +  0.0204i &   0.0308 -  0.0306i &  -0.0569 -  0.4699i \\
 -0.0622 -  0.0204i &   0.0135 &   0.0053 -  0.0015i &  -0.0050 +  0.0533i \\
  0.0308 +  0.0306i &   0.0053 +  0.0015i &   0.0043 &  -0.0039 -  0.0220i \\
 -0.0569 +  0.4699i &  -0.0050 -  0.0533i &  -0.0039 +  0.0220i &   0.4271 \\
\end{bmatrix}
$$

$$\rho_{O,9}=
\begin{bmatrix}
  0.0089 &  -0.0219 -  0.0731i &  -0.0027 +  0.0214i &  -0.0130 +  0.0040i \\
 -0.0219 +  0.0731i &   0.5684 &  -0.4492 +  0.0011i &  -0.0065 -  0.0551i \\
 -0.0027 -  0.0214i &  -0.4492 -  0.0011i &   0.4209 &  -0.0068 -  0.0044i \\
 -0.0130 -  0.0040i &  -0.0065 +  0.0551i &  -0.0068 +  0.0044i &   0.0018 \\
\end{bmatrix}
$$

$$\rho_{O,10}=
\begin{bmatrix}
  0.0058 &   0.0406 -  0.0232i &   0.0029 +  0.0025i &  -0.0009 +  0.0092i \\
  0.0406 +  0.0232i &   0.5133 &  -0.0158 +  0.4665i &  -0.0279 -  0.0088i \\
  0.0029 -  0.0025i &  -0.0158 -  0.4665i &   0.4753 &   0.0465 +  0.0408i \\
 -0.0009 -  0.0092i &  -0.0279 +  0.0088i &   0.0465 -  0.0408i &   0.0056 \\
\end{bmatrix}
$$

$$\rho_{O,11}=
\begin{bmatrix}
  0.2918 &   0.2395 -  0.0021i &  -0.0242 +  0.2258i &   0.0050 -  0.2663i \\
  0.2395 +  0.0021i &   0.1992 &  -0.0274 +  0.2259i &   0.0132 -  0.2382i \\
 -0.0242 -  0.2258i &  -0.0274 -  0.2259i &   0.2948 &  -0.2469 -  0.0169i \\
  0.0050 +  0.2663i &   0.0132 +  0.2382i &  -0.2469 +  0.0169i &   0.2142 \\
\end{bmatrix}
$$

$$\rho_{O,12}=
\begin{bmatrix}
  0.1923 &  -0.2541 -  0.0502i &  -0.0467 -  0.2470i &  -0.0163 -  0.2401i \\
 -0.2541 +  0.0502i &   0.3002 &   0.0111 +  0.2257i &   0.0075 +  0.2518i \\
 -0.0467 +  0.2470i &   0.0111 -  0.2257i &   0.2127 &   0.2396 +  0.0827i \\
 -0.0163 +  0.2401i &   0.0075 -  0.2518i &   0.2396 -  0.0827i &   0.2947 \\
\end{bmatrix}
$$

$$\rho_{O,13}=
\begin{bmatrix}
  0.3064 &  -0.2359 -  0.0153i &   0.0152 +  0.2270i &   0.0439 +  0.2918i \\
 -0.2359 +  0.0153i &   0.1884 &  -0.0182 -  0.1992i &  -0.0356 -  0.2367i \\
  0.0152 -  0.2270i &  -0.0182 +  0.1992i &   0.2840 &   0.2487 +  0.0009i \\
  0.0439 -  0.2918i &  -0.0356 +  0.2367i &   0.2487 -  0.0009i &   0.2211 \\
\end{bmatrix}
$$

$$\rho_{O,14}=
\begin{bmatrix}
  0.3803 &  -0.0814 -  0.3887i &  -0.1318 +  0.1051i &   0.1131 +  0.0665i \\
 -0.0814 +  0.3887i &   0.3748 &  -0.0973 -  0.1265i &  -0.0967 +  0.1000i \\
 -0.1318 -  0.1051i &  -0.0973 +  0.1265i &   0.1345 &  -0.0398 -  0.1196i \\
  0.1131 -  0.0665i &  -0.0967 -  0.1000i &  -0.0398 +  0.1196i &   0.1104 \\
\end{bmatrix}
$$

$$\rho_{O,15}=
\begin{bmatrix}
  0.5276 &   0.1680 +  0.4105i &  -0.0748 +  0.0527i &  -0.0341 -  0.0538i \\
  0.1680 -  0.4105i &   0.3819 &   0.0471 +  0.0874i &  -0.0793 +  0.0196i \\
 -0.0748 -  0.0527i &   0.0471 -  0.0874i &   0.0478 &   0.0094 +  0.0408i \\
 -0.0341 +  0.0538i &  -0.0793 -  0.0196i &   0.0094 -  0.0408i &   0.0427 \\
\end{bmatrix}
$$

$$\rho_{O,16}=
\begin{bmatrix}
  0.0207 &  -0.0261 -  0.1296i &  -0.0029 +  0.0401i &  -0.0233 -  0.0012i \\
 -0.0261 +  0.1296i &   0.6312 &  -0.3619 -  0.0096i &  -0.0148 -  0.0912i \\
 -0.0029 -  0.0401i &  -0.3619 +  0.0096i &   0.3450 &   0.0080 +  0.0213i \\
 -0.0233 +  0.0012i &  -0.0148 +  0.0912i &   0.0080 -  0.0213i &   0.0031 \\
\end{bmatrix}
$$

$$\rho_{O,17}=
\begin{bmatrix}
  0.0202 &  -0.0269 -  0.1385i &  -0.0033 +  0.0486i &  -0.0223 +  0.0155i \\
 -0.0269 +  0.1385i &   0.7415 &  -0.3405 +  0.0102i &  -0.0171 -  0.0985i \\
 -0.0033 -  0.0486i &  -0.3405 -  0.0102i &   0.2341 &   0.0106 +  0.0049i \\
 -0.0223 -  0.0155i &  -0.0171 +  0.0985i &   0.0106 -  0.0049i &   0.0042 \\
\end{bmatrix}
$$

\subsection{Pauli matrices based tomography}
$$\rho_{P,1}=
\begin{bmatrix}
  0.4879 &  -0.0194 +  0.0278i &   0.0045 +  0.0413i &  -0.4516 -  0.0735i \\
 -0.0194 -  0.0278i &   0.0054 &  -0.0053 +  0.0223i &   0.0336 +  0.0064i \\
  0.0045 -  0.0413i &  -0.0053 -  0.0223i &   0.0225 &  -0.0033 +  0.0768i \\
 -0.4516 +  0.0735i &   0.0336 -  0.0064i &  -0.0033 -  0.0768i &   0.4842 \\
\end{bmatrix}
$$
$$\rho_{P,2}=
\begin{bmatrix}
  0.4748 &   0.0191 -  0.0443i &   0.0193 -  0.0130i &   0.4662 +  0.0508i \\
  0.0191 +  0.0443i &   0.0107 &  -0.0000 -  0.0216i &   0.0200 +  0.0379i \\
  0.0193 +  0.0130i &  -0.0000 +  0.0216i &   0.0079 &   0.0111 -  0.0464i \\
  0.4662 -  0.0508i &   0.0200 -  0.0379i &   0.0111 +  0.0464i &   0.5065 \\
\end{bmatrix}
$$
$$\rho_{P,3}=
\begin{bmatrix}
  0.5221 &   0.0816 -  0.0442i &   0.0847 -  0.0240i &  -0.0026 +  0.4523i \\
  0.0816 +  0.0442i &   0.0213 &   0.0509 -  0.0035i &  -0.0463 +  0.0584i \\
  0.0847 +  0.0240i &   0.0509 +  0.0035i &   0.0112 &   0.0095 +  0.0545i \\
 -0.0026 -  0.4523i &  -0.0463 -  0.0584i &   0.0095 -  0.0545i &   0.4454 \\
\end{bmatrix}
$$
$$\rho_{P,4}=
\begin{bmatrix}
  0.2807 &   0.2569 +  0.0406i &   0.0456 -  0.2118i &  -0.0623 +  0.2607i \\
  0.2569 -  0.0406i &   0.2543 &   0.0401 -  0.1770i &  -0.0304 +  0.2331i \\
  0.0456 +  0.2118i &   0.0401 +  0.1770i &   0.1256 &  -0.2027 -  0.0279i \\
 -0.0623 -  0.2607i &  -0.0304 -  0.2331i &  -0.2027 +  0.0279i &   0.3394 \\
\end{bmatrix}
$$
$$\rho_{P,5}=
\begin{bmatrix}
  0.0260 &  -0.0969 +  0.0186i &   0.0380 +  0.0730i &   0.0156 +  0.0018i \\
 -0.0969 -  0.0186i &   0.5035 &   0.0389 -  0.4553i &   0.0051 -  0.0374i \\
  0.0380 -  0.0730i &   0.0389 +  0.4553i &   0.4676 &  -0.0051 -  0.0057i \\
  0.0156 -  0.0018i &   0.0051 +  0.0374i &  -0.0051 +  0.0057i &   0.0029 \\
\end{bmatrix}
$$
$$\rho_{P,6}=
\begin{bmatrix}
  0.0367 &   0.0502 +  0.0923i &   0.0734 +  0.0820i &  -0.0044 +  0.0326i \\
  0.0502 -  0.0923i &   0.4500 &   0.4397 -  0.0152i &  -0.0288 -  0.0230i \\
  0.0734 -  0.0820i &   0.4397 +  0.0152i &   0.5064 &  -0.0430 -  0.0023i \\
 -0.0044 -  0.0326i &  -0.0288 +  0.0230i &  -0.0430 +  0.0023i &   0.0069 \\
\end{bmatrix}
$$
$$\rho_{P,7}=
\begin{bmatrix}
  0.0062 &   0.0557 -  0.0422i &   0.0047 +  0.0036i &  -0.0086 -  0.0005i \\
  0.0557 +  0.0422i &   0.9818 &   0.0283 +  0.0662i &  -0.0441 -  0.1031i \\
  0.0047 -  0.0036i &   0.0283 -  0.0662i &   0.0032 &  -0.0049 +  0.0027i \\
 -0.0086 +  0.0005i &  -0.0441 +  0.1031i &  -0.0049 -  0.0027i &   0.0087 \\
\end{bmatrix}
$$
$$\rho_{P,8}=
\begin{bmatrix}
  0.5550 &  -0.0622 +  0.0204i &   0.0308 -  0.0306i &   0.0026 -  0.4514i \\
 -0.0622 -  0.0204i &   0.0135 &  -0.0451 +  0.0076i &  -0.0050 +  0.0533i \\
  0.0308 +  0.0306i &  -0.0451 -  0.0076i &   0.0043 &  -0.0039 -  0.0220i \\
  0.0026 +  0.4514i &  -0.0050 -  0.0533i &  -0.0039 +  0.0220i &   0.4271 \\
\end{bmatrix}
$$
$$\rho_{P,9}=
\begin{bmatrix}
  0.0089 &  -0.0219 -  0.0731i &  -0.0027 +  0.0214i &  -0.0215 -  0.0387i \\
 -0.0219 +  0.0731i &   0.5684 &  -0.4398 -  0.0689i &  -0.0065 -  0.0551i \\
 -0.0027 -  0.0214i &  -0.4398 +  0.0689i &   0.4209 &  -0.0068 -  0.0044i \\
 -0.0215 +  0.0387i &  -0.0065 +  0.0551i &  -0.0068 +  0.0044i &   0.0018 \\
\end{bmatrix}
$$
$$\rho_{P,10}=
\begin{bmatrix}
  0.0058 &   0.0406 -  0.0232i &   0.0029 +  0.0025i &  -0.0090 -  0.0064i \\
  0.0406 +  0.0232i &   0.5133 &   0.0500 +  0.4689i &  -0.0279 -  0.0088i \\
  0.0029 -  0.0025i &   0.0500 -  0.4689i &   0.4753 &   0.0465 +  0.0408i \\
 -0.0090 +  0.0064i &  -0.0279 +  0.0088i &   0.0465 -  0.0408i &   0.0056 \\
\end{bmatrix}
$$
$$\rho_{P,11}=
\begin{bmatrix}
  0.2918 &   0.2395 -  0.0021i &  -0.0242 +  0.2258i &   0.0052 -  0.2556i \\
  0.2395 +  0.0021i &   0.1992 &  -0.0291 +  0.2209i &   0.0132 -  0.2382i \\
 -0.0242 -  0.2258i &  -0.0291 -  0.2209i &   0.2948 &  -0.2469 -  0.0169i \\
  0.0052 +  0.2556i &   0.0132 +  0.2382i &  -0.2469 +  0.0169i &   0.2142 \\
\end{bmatrix}
$$
$$\rho_{P,12}=
\begin{bmatrix}
  0.1923 &  -0.2541 -  0.0502i &  -0.0467 -  0.2470i &   0.0350 -  0.2186i \\
 -0.2541 +  0.0502i &   0.3002 &   0.0458 +  0.2493i &   0.0075 +  0.2518i \\
 -0.0467 +  0.2470i &   0.0458 -  0.2493i &   0.2127 &   0.2396 +  0.0827i \\
  0.0350 +  0.2186i &   0.0075 -  0.2518i &   0.2396 -  0.0827i &   0.2947 \\
\end{bmatrix}
$$
$$\rho_{P,13}=
\begin{bmatrix}
  0.3064 &  -0.2359 -  0.0153i &   0.0152 +  0.2270i &   0.0170 +  0.2694i \\
 -0.2359 +  0.0153i &   0.1884 &   0.0105 -  0.2216i &  -0.0356 -  0.2367i \\
  0.0152 -  0.2270i &   0.0105 +  0.2216i &   0.2840 &   0.2487 +  0.0009i \\
  0.0170 -  0.2694i &  -0.0356 +  0.2367i &   0.2487 -  0.0009i &   0.2211 \\
\end{bmatrix}
$$
$$\rho_{P,14}=
\begin{bmatrix}
  0.3803 &  -0.0814 -  0.3887i &  -0.1318 +  0.1051i &   0.1286 +  0.0801i \\
 -0.0814 +  0.3887i &   0.3748 &  -0.0803 -  0.1471i &  -0.0967 +  0.1000i \\
 -0.1318 -  0.1051i &  -0.0803 +  0.1471i &   0.1345 &  -0.0398 -  0.1196i \\
  0.1286 -  0.0801i &  -0.0967 -  0.1000i &  -0.0398 +  0.1196i &   0.1104 \\
\end{bmatrix}
$$
$$\rho_{P,15}=
\begin{bmatrix}
  0.5276 &   0.1680 +  0.4105i &  -0.0748 +  0.0527i &  -0.0633 -  0.0764i \\
  0.1680 -  0.4105i &   0.3819 &   0.0050 +  0.0826i &  -0.0793 +  0.0196i \\
 -0.0748 -  0.0527i &   0.0050 -  0.0826i &   0.0478 &   0.0094 +  0.0408i \\
 -0.0633 +  0.0764i &  -0.0793 -  0.0196i &   0.0094 -  0.0408i &   0.0427 \\
\end{bmatrix}
$$
$$\rho_{P,16}=
\begin{bmatrix}
  0.0207 &  -0.0261 -  0.1296i &  -0.0029 +  0.0401i &  -0.0181 -  0.0374i \\
 -0.0261 +  0.1296i &   0.6312 &  -0.4367 -  0.0605i &  -0.0148 -  0.0912i \\
 -0.0029 -  0.0401i &  -0.4367 +  0.0605i &   0.3450 &   0.0080 +  0.0213i \\
 -0.0181 +  0.0374i &  -0.0148 +  0.0912i &   0.0080 -  0.0213i &   0.0031 \\
\end{bmatrix}
$$
$$\rho_{P,17}=
\begin{bmatrix}
  0.0202 &  -0.0269 -  0.1385i &  -0.0033 +  0.0486i &  -0.0181 -  0.0334i \\
 -0.0269 +  0.1385i &   0.7415 &  -0.3803 -  0.0547i &  -0.0171 -  0.0985i \\
 -0.0033 -  0.0486i &  -0.3803 +  0.0547i &   0.2341 &   0.0106 +  0.0049i \\
 -0.0181 +  0.0334i &  -0.0171 +  0.0985i &   0.0106 -  0.0049i &   0.0042 \\
\end{bmatrix}
$$

\section{Coefficient matrices}
All the analyzed tomographies  are based on solving the
linear-system problem
$$Ax = b,$$ where $A$ is the \emph{coefficient matrix}, ${b}$ is the
\emph{observation vector}, and $x={\rm vec}(\rho)$ is a real
vector describing the unknown state $\rho$ , i.e.,
$$
x={\rm vec}(\rho) = [\rho_{11},{\rm Re} \rho_{12},{\rm Im}
\rho_{12},{\rm Re} \rho_{13},{\rm Im} \rho_{13}, ...,\rho_{44}]^T.
$$
Thus, a two-qubit density matrix $\rho$ is represented as a real
vector $x=(x_1,...,x_{16})$ with its elements given as follows
$$
  \rho(x) = \left[
\begin{array}{cccc}
 x_{1} & x_{2}+i x_{3} & x_{4}+i x_{5} & x_{6}+i x_{7} \\
 x_{2}-i x_{3} & x_{8} & x_{9}+i x_{10} & x_{11}+i x_{12} \\
 x_{4}-i x_{5} & x_{9}-i x_{10} & x_{13} & x_{14}+i x_{15} \\
 x_{6}-i x_{7} & x_{11}-i x_{12} & x_{14}-i x_{15} & x_{16} \\
\end{array}
\right].
$$
The coefficient matrices depend on the choice of the equations
used for reconstructing a given density matrix. Below we list the
transposed (for typographic reasons) coefficient matrices for the
four analyzed tomographic protocols:

$$
A^{T}_{P} = \left[
\begin{array}{cccccccccccccccc}
0 & 0 & 0 & 0 & 0 & 2 & 0 & 0 & 2 & 0 & 0 & 0 & 0 & 0 & 0 & 0 \\
0 & 0 & 0 & 0 & 0 & 0 & -2 & 0 & 0 & 2 & 0 & 0 & 0 & 0 & 0 & 0 \\
0 & 0 & 0 & 2 & 0 & 0 & 0 & 0 & 0 & 0 & -2 & 0 & 0 & 0 & 0 & 0 \\
0 & 0 & 0 & 2 & 0 & 0 & 0 & 0 & 0 & 0 & 2 & 0 & 0 & 0 & 0 & 0 \\
0 & 0 & 0 & 0 & 0 & 0 & -2 & 0 & 0 & -2 & 0 & 0 & 0 & 0 & 0 & 0 \\
0 & 0 & 0 & 0 & 0 & -2 & 0 & 0 & 2 & 0 & 0 & 0 & 0 & 0 & 0 & 0 \\
0 & 0 & 0 & 0 & -2 & 0 & 0 & 0 & 0 & 0 & 0 & 2 & 0 & 0 & 0 & 0 \\
0 & 0 & 0 & 0 & -2 & 0 & 0 & 0 & 0 & 0 & 0 & -2 & 0 & 0 & 0 & 0 \\
0 & 2 & 0 & 0 & 0 & 0 & 0 & 0 & 0 & 0 & 0 & 0 & 0 & -2 & 0 & 0 \\
0 & 0 & -2 & 0 & 0 & 0 & 0 & 0 & 0 & 0 & 0 & 0 & 0 & 0 & 2 & 0 \\
1 & 0 & 0 & 0 & 0 & 0 & 0 & -1 & 0 & 0 & 0 & 0 & -1 & 0 & 0 & 1 \\
1 & 0 & 0 & 0 & 0 & 0 & 0 & 1 & 0 & 0 & 0 & 0 & -1 & 0 & 0 & -1 \\
0 & 2 & 0 & 0 & 0 & 0 & 0 & 0 & 0 & 0 & 0 & 0 & 0 & 2 & 0 & 0 \\
0 & 0 & -2 & 0 & 0 & 0 & 0 & 0 & 0 & 0 & 0 & 0 & 0 & 0 & -2 & 0 \\
1 & 0 & 0 & 0 & 0 & 0 & 0 & -1 & 0 & 0 & 0 & 0 & 1 & 0 & 0 & -1 \\
1 & 0 & 0 & 0 & 0 & 0 & 0 & 1 & 0 & 0 & 0 & 0 & 1 & 0 & 0 & 1 \\
\end{array}
\right]
$$

$$
A^{T}_{S} = \frac14\left[\begin{array}{ccccccccccccccccc}
 4 &  0 &  0 &  0 &  0 &  0 &  0 &  0 &  0 &  0 &  0 &  0 &  0 &  0 &  0 &  0 \\
 0 &  0 &  0 &  0 &  0 &  0 &  0 &  4 &  0 &  0 &  0 &  0 &  0 &  0 &  0 &  0 \\
 2 &  4 &  0 &  0 &  0 &  0 &  0 &  2 &  0 &  0 &  0 &  0 &  0 &  0 &  0 &  0 \\
 2 &  -4 &  0 &  0 &  0 &  0 &  0 &  2 &  0 &  0 &  0 &  0 &  0 &  0 &  0 &  0 \\
 2 &  0 &  -4 &  0 &  0 &  0 &  0 &  2 &  0 &  0 &  0 &  0 &  0 &  0 &  0 &  0 \\
 2 &  0 &  4 &  0 &  0 &  0 &  0 &  2 &  0 &  0 &  0 &  0 &  0 &  0 &  0 &  0 \\
 0 &  0 &  0 &  0 &  0 &  0 &  0 &  0 &  0 &  0 &  0 &  0 &  4 &  0 &  0 &  0 \\
 0 &  0 &  0 &  0 &  0 &  0 &  0 &  0 &  0 &  0 &  0 &  0 &  0 &  0 &  0 &  4 \\
 0 &  0 &  0 &  0 &  0 &  0 &  0 &  0 &  0 &  0 &  0 &  0 &  2 &  4 &  0 &  2 \\
 0 &  0 &  0 &  0 &  0 &  0 &  0 &  0 &  0 &  0 &  0 &  0 &  2 &  -4 &  0 &  2 \\
 0 &  0 &  0 &  0 &  0 &  0 &  0 &  0 &  0 &  0 &  0 &  0 &  2 &  0 &  -4 &  2 \\
 0 &  0 &  0 &  0 &  0 &  0 &  0 &  0 &  0 &  0 &  0 &  0 &  2 &  0 &  4 &  2 \\
 2 &  0 &  0 &  4 &  0 &  0 &  0 &  0 &  0 &  0 &  0 &  0 &  2 &  0 &  0 &  0 \\
 0 &  0 &  0 &  0 &  0 &  0 &  0 &  2 &  0 &  0 &  4 &  0 &  0 &  0 &  0 &  2 \\
 1 &  2 &  0 &  2 &  0 &  2 &  0 &  1 &  2 &  0 &  2 &  0 &  1 &  2 &  0 &  1 \\
 1 &  -2 &  0 &  2 &  0 &  -2 &  0 &  1 &  -2 &  0 &  2 &  0 &  1 &  -2 &  0 &  1
\\
 1 &  0 &  -2 &  2 &  0 &  0 &  -2 &  1 &  0 &  2 &  2 &  0 &  1 &  0 &  -2 &  1
\\
 1 &  0 &  2 &  2 &  0 &  0 &  2 &  1 &  0 &  -2 &  2 &  0 &  1 &  0 &  2 &  1 \\
 2 &  0 &  0 &  -4 &  0 &  0 &  0 &  0 &  0 &  0 &  0 &  0 &  2 &  0 &  0 &  0 \\
 0 &  0 &  0 &  0 &  0 &  0 &  0 &  2 &  0 &  0 &  -4 &  0 &  0 &  0 &  0 &  2 \\
 1 &  2 &  0 &  -2 &  0 &  -2 &  0 &  1 &  -2 &  0 &  -2 &  0 &  1 &  2 &  0 &  1
\\
 1 &  -2 &  0 &  -2 &  0 &  2 &  0 &  1 &  2 &  0 &  -2 &  0 &  1 &  -2 &  0 &  1
\\
 1 &  0 &  -2 &  -2 &  0 &  0 &  2 &  1 &  0 &  -2 &  -2 &  0 &  1 &  0 &  -2 &
1 \\
 1 &  0 &  2 &  -2 &  0 &  0 &  -2 &  1 &  0 &  2 &  -2 &  0 &  1 &  0 &  2 &  1
\\
 2 &  0 &  0 &  0 &  -4 &  0 &  0 &  0 &  0 &  0 &  0 &  0 &  2 &  0 &  0 &  0 \\
 0 &  0 &  0 &  0 &  0 &  0 &  0 &  2 &  0 &  0 &  0 &  -4 &  0 &  0 &  0 &  2 \\
 1 &  2 &  0 &  0 &  -2 &  0 &  -2 &  1 &  0 &  -2 &  0 &  -2 &  1 &  2 &  0 &  1
\\
 1 &  -2 &  0 &  0 &  -2 &  0 &  2 &  1 &  0 &  2 &  0 &  -2 &  1 &  -2 &  0 &  1
\\
 1 &  0 &  -2 &  0 &  -2 &  -2 &  0 &  1 &  2 &  0 &  0 &  -2 &  1 &  0 &  -2 &
1 \\
 1 &  0 &  2 &  0 &  -2 &  2 &  0 &  1 &  -2 &  0 &  0 &  -2 &  1 &  0 &  2 &  1
\\
 2 &  0 &  0 &  0 &  4 &  0 &  0 &  0 &  0 &  0 &  0 &  0 &  2 &  0 &  0 &  0 \\
 0 &  0 &  0 &  0 &  0 &  0 &  0 &  2 &  0 &  0 &  0 &  4 &  0 &  0 &  0 &  2 \\
 1 &  2 &  0 &  0 &  2 &  0 &  2 &  1 &  0 &  2 &  0 &  2 &  1 &  2 &  0 &  1 \\
 1 &  -2 &  0 &  0 &  2 &  0 &  -2 &  1 &  0 &  -2 &  0 &  2 &  1 &  -2 &  0 &  1
\\
 1 &  0 &  -2 &  0 &  2 &  2 &  0 &  1 &  -2 &  0 &  0 &  2 &  1 &  0 &  -2 &  1
\\
 1 &  0 &  2 &  0 &  2 &  -2 &  0 &  1 &  2 &  0 &  0 &  2 &  1 &  0 &  2 &  1 \\
 \end{array}
\right]
$$

$$
A^{T}_{J} = \frac14\left[\begin{array}{ccccccccccccccccc}
 4 &  0 &  0 &  0 &  0 &  0 &  0 &  0 &  0 &  0 &  0 &  0 &  0 &  0 &  0 &  0 \\
 0 &  0 &  0 &  0 &  0 &  0 &  0 &  4 &  0 &  0 &  0 &  0 &  0 &  0 &  0 &  0 \\
 2 &  4 &  0 &  0 &  0 &  0 &  0 &  2 &  0 &  0 &  0 &  0 &  0 &  0 &  0 &  0 \\
 2 &  0 &  -4 &  0 &  0 &  0 &  0 &  2 &  0 &  0 &  0 &  0 &  0 &  0 &  0 &  0 \\
 0 &  0 &  0 &  0 &  0 &  0 &  0 &  0 &  0 &  0 &  0 &  0 &  4 &  0 &  0 &  0 \\
 0 &  0 &  0 &  0 &  0 &  0 &  0 &  0 &  0 &  0 &  0 &  0 &  0 &  0 &  0 &  4 \\
 0 &  0 &  0 &  0 &  0 &  0 &  0 &  0 &  0 &  0 &  0 &  0 &  2 &  4 &  0 &  2 \\
 0 &  0 &  0 &  0 &  0 &  0 &  0 &  0 &  0 &  0 &  0 &  0 &  2 &  0 &  -4 &  2 \\
 2 &  0 &  0 &  0 &  4 &  0 &  0 &  0 &  0 &  0 &  0 &  0 &  2 &  0 &  0 &  0 \\
 0 &  0 &  0 &  0 &  0 &  0 &  0 &  2 &  0 &  0 &  0 &  4 &  0 &  0 &  0 &  2 \\
 1 &  2 &  0 &  0 &  2 &  0 &  2 &  1 &  0 &  2 &  0 &  2 &  1 &  2 &  0 &  1 \\
 1 &  0 &  -2 &  0 &  2 &  2 &  0 &  1 &  -2 &  0 &  0 &  2 &  1 &  0 &  -2 &  1
\\
 2 &  0 &  0 &  4 &  0 &  0 &  0 &  0 &  0 &  0 &  0 &  0 &  2 &  0 &  0 &  0 \\
 0 &  0 &  0 &  0 &  0 &  0 &  0 &  2 &  0 &  0 &  4 &  0 &  0 &  0 &  0 &  2 \\
 1 &  2 &  0 &  2 &  0 &  2 &  0 &  1 &  2 &  0 &  2 &  0 &  1 &  2 &  0 &  1 \\
 1 &  0 &  2 &  2 &  0 &  0 &  2 &  1 &  0 &  -2 &  2 &  0 &  1 &  0 &  2 &  1 \\
 \end{array}
\right]
$$

$$
A^{T}_{M} =\frac14 \left[\begin{array}{ccccccccccccccccc}
 2 &  0 &  0 &  4 &  0 &  0 &  0 &  0 &  0 &  0 &  0 &  0 &  2 &  0 &  0 &  0 \\
 0 &  0 &  0 &  0 &  0 &  0 &  0 &  2 &  0 &  0 &  4 &  0 &  0 &  0 &  0 &  2 \\
 2 &  0 &  0 &  -4 &  0 &  0 &  0 &  0 &  0 &  0 &  0 &  0 &  2 &  0 &  0 &  0 \\
 0 &  0 &  0 &  0 &  0 &  0 &  0 &  2 &  0 &  0 &  -4 &  0 &  0 &  0 &  0 &  2 \\
 1 &  2 &  0 &  0 &  -2 &  0 &  -2 &  1 &  0 &  -2 &  0 &  -2 &  1 &  2 &  0 &  1
\\
 1 &  -2 &  0 &  0 &  -2 &  0 &  2 &  1 &  0 &  2 &  0 &  -2 &  1 &  -2 &  0 &  1
\\
 1 &  2 &  0 &  0 &  2 &  0 &  2 &  1 &  0 &  2 &  0 &  2 &  1 &  2 &  0 &  1 \\
 1 &  -2 &  0 &  0 &  2 &  0 &  -2 &  1 &  0 &  -2 &  0 &  2 &  1 &  -2 &  0 &  1
\\
 0 &  0 &  0 &  0 &  0 &  0 &  0 &  0 &  0 &  0 &  0 &  0 &  2 &  0 &  4 &  2 \\
 0 &  0 &  0 &  0 &  0 &  0 &  0 &  0 &  0 &  0 &  0 &  0 &  2 &  0 &  -4 &  2 \\
 2 &  0 &  4 &  0 &  0 &  0 &  0 &  2 &  0 &  0 &  0 &  0 &  0 &  0 &  0 &  0 \\
 2 &  0 &  -4 &  0 &  0 &  0 &  0 &  2 &  0 &  0 &  0 &  0 &  0 &  0 &  0 &  0 \\
 2 &  0 &  0 &  0 &  0 &  4 &  0 &  0 &  0 &  0 &  0 &  0 &  0 &  0 &  0 &  2 \\
 2 &  0 &  0 &  0 &  0 &  -4 &  0 &  0 &  0 &  0 &  0 &  0 &  0 &  0 &  0 &  2 \\
 0 &  0 &  0 &  0 &  0 &  0 &  0 &  2 &  4 &  0 &  0 &  0 &  2 &  0 &  0 &  0 \\
 0 &  0 &  0 &  0 &  0 &  0 &  0 &  2 &  -4 &  0 &  0 &  0 &  2 &  0 &  0 &  0 \\
 1 &  2 &  0 &  0 &  2 &  0 &  -2 &  1 &  0 &  2 &  0 &  -2 &  1 &  -2 &  0 &  1
\\
 1 &  -2 &  0 &  0 &  -2 &  0 &  -2 &  1 &  0 &  2 &  0 &  2 &  1 &  2 &  0 &  1
\\
 1 &  -2 &  0 &  0 &  2 &  0 &  2 &  1 &  0 &  -2 &  0 &  -2 &  1 &  2 &  0 &  1
\\
 1 &  2 &  0 &  0 &  -2 &  0 &  2 &  1 &  0 &  -2 &  0 &  2 &  1 &  -2 &  0 &  1
\\
 \end{array}
\right]
$$

$$
A^{T}_{O} = \left[\begin{array}{ccccccccccccccccc}
 1 &  0 &  0 &  0 &  0 &  0 &  0 &  0 &  0 &  0 &  0 &  0 &  0 &  0 &  0 &  0 \\
 0 &  0 &  0 &  0 &  0 &  0 &  0 &  1 &  0 &  0 &  0 &  0 &  0 &  0 &  0 &  0 \\
 0 &  0 &  0 &  0 &  0 &  0 &  0 &  0 &  0 &  0 &  0 &  0 &  1 &  0 &  0 &  0 \\
 0 &  0 &  0 &  0 &  0 &  0 &  0 &  0 &  0 &  0 &  0 &  0 &  0 &  0 &  0 &  1 \\
 0 &  1 &  0 &  0 &  0 &  0 &  0 &  0 &  0 &  0 &  0 &  0 &  0 &  0 &  0 &  0 \\
 0 &  0 &  -1 &  0 &  0 &  0 &  0 &  0 &  0 &  0 &  0 &  0 &  0 &  0 &  0 &  0 \\
 0 &  0 &  0 &  1 &  0 &  0 &  0 &  0 &  0 &  0 &  0 &  0 &  0 &  0 &  0 &  0 \\
 0 &  0 &  0 &  0 &  -1 &  0 &  0 &  0 &  0 &  0 &  0 &  0 &  0 &  0 &  0 &  0 \\
 0 &  0 &  0 &  0 &  0 &  0 &  0 &  0 &  0 &  0 &  0 &  0 &  0 &  1 &  0 &  0 \\
 0 &  0 &  0 &  0 &  0 &  0 &  0 &  0 &  0 &  0 &  0 &  0 &  0 &  0 &  -1 &  0 \\
 0 &  0 &  0 &  0 &  0 &  0 &  0 &  0 &  0 &  0 &  1 &  0 &  0 &  0 &  0 &  0 \\
 0 &  0 &  0 &  0 &  0 &  0 &  0 &  0 &  0 &  0 &  0 &  -1 &  0 &  0 &  0 &  0 \\
 0 &  0 &  0 &  0 &  0 &  0 &  0 &  0 &  1 &  0 &  0 &  0 &  0 &  0 &  0 &  0 \\
 0 &  0 &  0 &  0 &  0 &  0 &  0 &  0 &  0 &  -1 &  0 &  0 &  0 &  0 &  0 &  0 \\
 0 &  0 &  0 &  0 &  0 &  1 &  0 &  0 &  0 &  0 &  0 &  0 &  0 &  0 &  0 &  0 \\
 0 &  0 &  0 &  0 &  0 &  0 &  -1 &  0 &  0 &  0 &  0 &  0 &  0 &  0 &  0 &  0 \\
 \end{array}
\right]
$$

\section{Observation vectors}
The observation vectors correspond to photon coincidence counts.
In reality we measure disturbed quantities $\bar b\equiv
b+\delta\,b$ instead of $b$. The observation vectors are column
vectors. For convenience we arrange them in arrays, where each
column corresponds to one of the 17 reconstructed states.
$$
\bar b = \bbordermatrix{ &\rho_1 & \rho_2 & \dots & \rho_{17}\cr
 b_1 & \bar b_{1,1}  & \bar b_{1,2} & \ldots & \bar b_{1,17}\cr
 b_2 & \bar b_{2,1} & \bar b_{2,2} & \ldots & \bar b_{2,17}\cr
 \vdots & \vdots   & \vdots  & \ddots & \vdots \cr
 b_N & \bar b_{N,1}& \bar b_{N,2}  & \ldots  & \bar b_{N,17} }
$$
Note that the values of $b$ listed below are not normalized and
cannot be interpreted as probabilities. The elements of each
vector $b$ were registered over 5 seconds. This means that if an
element of $b$ is a sum or a difference of $n$ projectors the
measurement for each of the $n$ projectors took $5/n$ seconds. In
this way the measurements for observation vectors of the same
length take the same amount of time. To obtain the frequencies we
can divide these values by the total number of photon coincidences
counted or by a sum of coincidences counted for a set of
projectors forming a basis. The set of such projectors is not
unique. In our calculations we use the unnormalized coincidences
and normalize the reconstructed density matrices.

\subsection{Standard 36 state tomography}

$$
\bar b_S = \left[\begin{array}{ccccccccccccccccc}
 2727 &  2575 &  2844 &  1448 &  127 &  193 &  25 &  2955 &  40 &  27 &  1264 &
809 &  1231 &  2762 &  3831 &  113 &  112 \\
 30 &  58 &  116 &  1312 &  2457 &  2364 &  3928 &  72 &  2555 &  2375 &  863 &
1263 &  757 &  2722 &  2773 &  3452 &  4102 \\
 1244 &  1401 &  1890 &  2688 &  826 &  1483 &  2159 &  1096 &  1152 &  1386 &  2126
&  40 &  50 &  2251 &  4529 &  1758 &  2012 \\
 1461 &  1194 &  1001 &  37 &  1772 &  956 &  1713 &  1758 &  1349 &  1010 &  51
&  2178 &  1945 &  3434 &  2090 &  2043 &  2310 \\
 1270 &  1697 &  1865 &  1224 &  1120 &  722 &  2102 &  1407 &  1433 &  1210 &  1150
&  1200 &  1057 &  5667 &  266 &  2409 &  2782 \\
 1581 &  1216 &  1383 &  1643 &  1302 &  1692 &  1764 &  1624 &  776 &  995 &  1132
&  778 &  934 &  21 &  6227 &  991 &  1250 \\
 126 &  43 &  61 &  648 &  2282 &  2660 &  13 &  23 &  1892 &  2199 &  1277 &  895
&  1141 &  977 &  347 &  1887 &  1295 \\
 2706 &  2747 &  2426 &  1751 &  14 &  36 &  35 &  2274 &  8 &  26 &  928 &  1240
&  888 &  802 &  310 &  17 &  23 \\
 1403 &  1448 &  1248 &  113 &  1051 &  1095 &  10 &  1134 &  951 &  1399 &  35 &
 2079 &  2034 &  646 &  352 &  974 &  719 \\
 1440 &  1328 &  1144 &  2204 &  1101 &  1547 &  49 &  1175 &  1012 &  969 &  2174
&  63 &  36 &  1224 &  216 &  887 &  602 \\
 940 &  1611 &  907 &  1238 &  1253 &  1422 &  11 &  1235 &  1018 &  1069 &  1099
&  756 &  1067 &  1769 &  9 &  852 &  664 \\
 1798 &  1108 &  1501 &  950 &  1197 &  1398 &  33 &  1001 &  978 &  1447 &  953
&  1452 &  1074 &  31 &  601 &  1085 &  718 \\
 1316 &  1304 &  1837 &  1250 &  1385 &  1781 &  42 &  1458 &  952 &  1172 &  1147
&  710 &  1251 &  901 &  1634 &  939 &  651 \\
 1607 &  1537 &  981 &  1279 &  1149 &  920 &  1594 &  1156 &  1207 &  1065 &  991
&  1268 &  691 &  1085 &  987 &  1753 &  1986 \\
 42 &  2845 &  1706 &  1231 &  1225 &  2556 &  962 &  1090 &  32 &  1482 &  996 &
 1111 &  1012 &  790 &  1732 &  81 &  208 \\
 2740 &  40 &  967 &  1276 &  1273 &  87 &  684 &  1568 &  2169 &  886 &  1160 &
 819 &  880 &  1359 &  755 &  2663 &  2424 \\
 1354 &  1497 &  107 &  25 &  25 &  1165 &  1024 &  2671 &  1050 &  2240 &  2215
&  1949 &  64 &  2083 &  131 &  1545 &  1485 \\
 1298 &  1459 &  2689 &  2352 &  2356 &  1818 &  564 &  43 &  876 &  156 &  47 &
 72 &  1866 &  5 &  2458 &  1053 &  938 \\
 1266 &  1095 &  914 &  780 &  1014 &  1010 &  4 &  1130 &  976 &  1145 &  1357 &
 1103 &  1129 &  2815 &  2720 &  971 &  688 \\
 1231 &  1320 &  1485 &  1593 &  1099 &  1223 &  1947 &  1209 &  1265 &  1323 &
877 &  1205 &  977 &  2489 &  2139 &  1915 &  2175 \\
 2459 &  57 &  1288 &  1250 &  791 &  28 &  1043 &  1137 &  2090 &  1251 &  1081
&  944 &  1118 &  2145 &  3266 &  2585 &  2534 \\
 50 &  2308 &  1075 &  1066 &  1371 &  2133 &  923 &  1163 &  80 &  1034 &  1038
&  1332 &  1207 &  3415 &  1442 &  192 &  342 \\
 818 &  1591 &  2448 &  2275 &  2158 &  947 &  875 &  22 &  1305 &  22 &  39 &  31
&  2170 &  5417 &  141 &  1841 &  1860 \\
 1833 &  767 &  65 &  86 &  28 &  1098 &  949 &  2281 &  859 &  2336 &  1999 &  2091
&  28 &  38 &  4777 &  1096 &  1077 \\
 1229 &  1498 &  1585 &  2277 &  712 &  772 &  2 &  1617 &  689 &  979 &  80 &  2142
&  61 &  1173 &  1991 &  601 &  314 \\
 1259 &  1048 &  878 &  158 &  1535 &  1390 &  2521 &  804 &  1576 &  1348 &  2087
&  66 &  1988 &  1094 &  1532 &  2478 &  2874 \\
 1743 &  924 &  123 &  1133 &  2117 &  918 &  1288 &  2314 &  1521 &  60 &  1199
&  856 &  799 &  994 &  2446 &  1994 &  2063 \\
 850 &  1559 &  2345 &  1188 &  213 &  1343 &  1326 &  81 &  703 &  2252 &  866 &
 1327 &  1284 &  1313 &  981 &  999 &  1180 \\
 2423 &  84 &  1608 &  1713 &  1499 &  2127 &  1416 &  936 &  189 &  970 &  718 &
 914 &  1206 &  2181 &  142 &  455 &  725 \\
 199 &  2354 &  828 &  733 &  902 &  34 &  1028 &  1672 &  1919 &  1103 &  1253 &
 1332 &  677 &  12 &  3277 &  2470 &  2414 \\
 1691 &  1357 &  1323 &  92 &  1424 &  1634 &  31 &  1291 &  881 &  1002 &  2036
&  64 &  1885 &  2699 &  2757 &  1040 &  852 \\
 1331 &  1459 &  1514 &  2563 &  1170 &  1148 &  1696 &  1372 &  1081 &  1267 &
23 &  2185 &  86 &  2546 &  1817 &  1481 &  1784 \\
 1648 &  1243 &  2746 &  1707 &  30 &  1217 &  1114 &  42 &  991 &  2140 &  1068
&  946 &  939 &  2106 &  2968 &  1278 &  1244 \\
 1327 &  1561 &  79 &  899 &  2552 &  1460 &  626 &  2535 &  1140 &  52 &  1036 &
 1158 &  1040 &  3398 &  1413 &  1354 &  1335 \\
 48 &  2814 &  1524 &  1198 &  1435 &  84 &  889 &  1204 &  2062 &  804 &  913 &
 746 &  1349 &  5230 &  157 &  2610 &  2403 \\
 2812 &  27 &  1327 &  1275 &  1066 &  2657 &  796 &  1432 &  32 &  1483 &  1151
&  1255 &  768 &  27 &  4284 &  47 &  85 \\
\end{array}
\right]
$$
The rows of the observation vector $\bar{b}_S$ correspond to the
following consecutive projectors: $\oprod{HH}$, $\oprod{HV}$,
$\oprod{HD}$, $\oprod{HA}$, $\oprod{HL}$, $\oprod{HR}$,
$\oprod{VH}$, $\oprod{VV}$, $\oprod{VD}$, $\oprod{VA}$,
$\oprod{VL}$, $\oprod{VR}$, $\oprod{DH}$, $\oprod{DV}$,
$\oprod{DD}$, $\oprod{DA}$, $\oprod{DL}$, $\oprod{DR}$,
$\oprod{AH}$, $\oprod{AV}$, $\oprod{AD}$, $\oprod{AA}$,
$\oprod{AL}$, $\oprod{AR}$, $\oprod{LH}$, $\oprod{LV}$,
$\oprod{LD}$, $\oprod{LA}$, $\oprod{LL}$, $\oprod{LR}$,
$\oprod{RH}$, $\oprod{RV}$, $\oprod{RD}$, $\oprod{RA}$,
$\oprod{RL}$, $\oprod{RR}$.

\subsection{ JKMW tomography}
$$
\bar b_J = \left[\begin{array}{ccccccccccccccccc}
 2727 &  2575 &  2844 &  1448 &  127 &  193 &  25 &  2955 &  40 &  27 &  1264 &
809 &  1231 &  2762 &  3831 &  113 &  112 \\
 30 &  58 &  116 &  1312 &  2457 &  2364 &  3928 &  72 &  2555 &  2375 &  863 &
1263 &  757 &  2722 &  2773 &  3452 &  4102 \\
 1244 &  1401 &  1890 &  2688 &  826 &  1483 &  2159 &  1096 &  1152 &  1386 &  2126
&  40 &  50 &  2251 &  4529 &  1758 &  2012 \\
 1270 &  1697 &  1865 &  1224 &  1120 &  722 &  2102 &  1407 &  1433 &  1210 &  1150
&  1200 &  1057 &  5667 &  266 &  2409 &  2782 \\
 126 &  43 &  61 &  648 &  2282 &  2660 &  13 &  23 &  1892 &  2199 &  1277 &  895
&  1141 &  977 &  347 &  1887 &  1295 \\
 2706 &  2747 &  2426 &  1751 &  14 &  36 &  35 &  2274 &  8 &  26 &  928 &  1240
&  888 &  802 &  310 &  17 &  23 \\
 1403 &  1448 &  1248 &  113 &  1051 &  1095 &  10 &  1134 &  951 &  1399 &  35 &
 2079 &  2034 &  646 &  352 &  974 &  719 \\
 940 &  1611 &  907 &  1238 &  1253 &  1422 &  11 &  1235 &  1018 &  1069 &  1099
&  756 &  1067 &  1769 &  9 &  852 &  664 \\
 1691 &  1357 &  1323 &  92 &  1424 &  1634 &  31 &  1291 &  881 &  1002 &  2036
&  64 &  1885 &  2699 &  2757 &  1040 &  852 \\
 1331 &  1459 &  1514 &  2563 &  1170 &  1148 &  1696 &  1372 &  1081 &  1267 &
23 &  2185 &  86 &  2546 &  1817 &  1481 &  1784 \\
 1648 &  1243 &  2746 &  1707 &  30 &  1217 &  1114 &  42 &  991 &  2140 &  1068
&  946 &  939 &  2106 &  2968 &  1278 &  1244 \\
 48 &  2814 &  1524 &  1198 &  1435 &  84 &  889 &  1204 &  2062 &  804 &  913 &
 746 &  1349 &  5230 &  157 &  2610 &  2403 \\
 1316 &  1304 &  1837 &  1250 &  1385 &  1781 &  42 &  1458 &  952 &  1172 &  1147
&  710 &  1251 &  901 &  1634 &  939 &  651 \\
 1607 &  1537 &  981 &  1279 &  1149 &  920 &  1594 &  1156 &  1207 &  1065 &  991
&  1268 &  691 &  1085 &  987 &  1753 &  1986 \\
 42 &  2845 &  1706 &  1231 &  1225 &  2556 &  962 &  1090 &  32 &  1482 &  996 &
 1111 &  1012 &  790 &  1732 &  81 &  208 \\
 1298 &  1459 &  2689 &  2352 &  2356 &  1818 &  564 &  43 &  876 &  156 &  47 &
 72 &  1866 &  5 &  2458 &  1053 &  938 \\
\end{array}
\right]
$$
The rows of the observation vector $\bar{b}_J$ correspond to the
following consecutive projectors: $\oprod{HH}$, $\oprod{HV}$,
$\oprod{HD}$, $\oprod{HL}$, $\oprod{VH}$, $\oprod{VV}$,
$\oprod{VD}$, $\oprod{VL}$, $\oprod{RH}$, $\oprod{RV}$,
$\oprod{RD}$, $\oprod{RL}$, $\oprod{DH}$, $\oprod{DV}$,
$\oprod{DD}$, $\oprod{DR}$.

\subsection{MUB-based tomography}

$$
\bar b_{M} = \left[\begin{array}{ccccccccccccccccc}
 1316 &  1304 &  1837 &  1250 &  1385 &  1781 &  42 &  1458 &  952 &  1172 &  1147
&  710 &  1251 &  901 &  1634 &  939 &  651 \\
 1607 &  1537 &  981 &  1279 &  1149 &  920 &  1594 &  1156 &  1207 &  1065 &  991
&  1268 &  691 &  1085 &  987 &  1753 &  1986 \\
 1266 &  1095 &  914 &  780 &  1014 &  1010 &  4 &  1130 &  976 &  1145 &  1357 &
 1103 &  1129 &  2815 &  2720 &  971 &  688 \\
 1231 &  1320 &  1485 &  1593 &  1099 &  1223 &  1947 &  1209 &  1265 &  1323 &
877 &  1205 &  977 &  2489 &  2139 &  1915 &  2175 \\
 1743 &  924 &  123 &  1133 &  2117 &  918 &  1288 &  2314 &  1521 &  60 &  1199
&  856 &  799 &  994 &  2446 &  1994 &  2063 \\
 850 &  1559 &  2345 &  1188 &  213 &  1343 &  1326 &  81 &  703 &  2252 &  866 &
 1327 &  1284 &  1313 &  981 &  999 &  1180 \\
 1648 &  1243 &  2746 &  1707 &  30 &  1217 &  1114 &  42 &  991 &  2140 &  1068
&  946 &  939 &  2106 &  2968 &  1278 &  1244 \\
 1327 &  1561 &  79 &  899 &  2552 &  1460 &  626 &  2535 &  1140 &  52 &  1036 &
 1158 &  1040 &  3398 &  1413 &  1354 &  1335 \\
 1798 &  1108 &  1501 &  950 &  1197 &  1398 &  33 &  1001 &  978 &  1447 &  953
&  1452 &  1074 &  31 &  601 &  1085 &  718 \\
 940 &  1611 &  907 &  1238 &  1253 &  1422 &  11 &  1235 &  1018 &  1069 &  1099
&  756 &  1067 &  1769 &  9 &  852 &  664 \\
 1581 &  1216 &  1383 &  1643 &  1302 &  1692 &  1764 &  1624 &  776 &  995 &  1132
&  778 &  934 &  21 &  6227 &  991 &  1250 \\
 1270 &  1697 &  1865 &  1224 &  1120 &  722 &  2102 &  1407 &  1433 &  1210 &  1150
&  1200 &  1057 &  5667 &  266 &  2409 &  2782 \\
 97 &  5284 &  2720 &  1181 &  135 &  134 &  75 &  2337 &  16 &  23 &  1323 &  958
&  1438 &  2600 &  1867 &  23 &  21 \\
 5418 &  98 &  2143 &  874 &  119 &  27 &  133 &  2943 &  133 &  31 &  1279 &  1095
&  1085 &  957 &  2362 &  277 &  268 \\
 32 &  55 &  87 &  1558 &  3494 &  4939 &  1892 &  88 &  175 &  2121 &  1091 &  868
&  1002 &  826 &  1537 &  586 &  623 \\
 157 &  27 &  109 &  478 &  1725 &  253 &  1887 &  32 &  4213 &  2268 &  1328 &
775 &  1148 &  2240 &  853 &  4544 &  4390 \\
 2022 &  1093 &  184 &  120 &  182 &  1636 &  1011 &  2320 &  786 &  1912 &  4230
&  57 &  48 &  607 &  2921 &  944 &  945 \\
 1245 &  851 &  111 &  130 &  41 &  646 &  637 &  2232 &  936 &  2583 &  52 &  4028
&  9 &  693 &  1795 &  931 &  998 \\
 1404 &  1117 &  1788 &  40 &  2693 &  1246 &  990 &  60 &  1415 &  71 &  43 &  70
&  3875 &  2676 &  868 &  1920 &  1922 \\
 1355 &  1652 &  3361 &  4078 &  1349 &  1156 &  613 &  115 &  862 &  81 &  13 &
 84 &  54 &  2240 &  1613 &  868 &  921 \\
\end{array}
\right]
$$
The rows of the observation vector $\bar{b}_M$ correspond to the
following consecutive projectors: $\oprod{DH}$, $\oprod{DV}$,
$\oprod{AH}$, $\oprod{AV}$, $\oprod{LD}$, $\oprod{LA}$,
$\oprod{RD}$, $\oprod{RA}$, $\oprod{VR}$, $\oprod{VL}$,
$\oprod{HR}$, $\oprod{HL}$, $\oprod{\Phi^+}$, $\oprod{\Phi^-}$,
$\oprod{\Psi^+}$,  $\oprod{\Psi^-}$, $\tfrac{1}{2}(\ket{DL} +
i\ket{AR})(\bra{DL} - i\bra{AR})$, $\tfrac{1}{2}(\ket{DL} -
i\ket{AR})(\bra{DL} + i\bra{AR})$, $\tfrac{1}{2}(\ket{DR} +
i\ket{AL})(\bra{DR} - i\bra{AL})$, $\tfrac{1}{2}(\ket{DR} -
i\ket{AL})(\bra{DR} + i\bra{AL})$, where
$\ket{\Phi^\pm}=(\ket{HH}\pm
 \ket{VV})/\sqrt{2}$ and  $\ket{ \Psi^\pm}=(\ket{HV}\pm
\ket{VH})/\sqrt{2}$.

\subsection{Optimal tomography}

$$
\bar b_{O} = \left[\begin{array}{ccccccccccccccccc} {2727} &
{2575} & {2844} & {1448} & {127} & {193} & {25} & {2955} & {40} &
{27} &
{1264} & {809} & {1231} & {2762} & {3831} & {113} & {112} \\
{30} & {58} & {116} & {1312} & {2457} & {2364} & {3928} & {72} &
{2555} & {2375}
& {863} & {1263} & {757} & {2722} & {2773} & {3452} & {4102} \\
{126} & {43} & {61} & {648} & {2282} & {2660} & {13} & {23} &
{1892} & {2199} & {1277}
& {895} & {1141} & {977} & {347} & {1887} & {1295} \\
{2706} & {2747} & {2426} & {1751} & {14} & {36} & {35} & {2274} &
{8} & {26} & {928}
& {1240} & {888} & {802} & {310} & {17} & {23} \\
\overline{108} & {103} & {444} & {1325} & \overline{473} & {263} &
{223} & \overline{331} & \overline{98} & {188} & {1037} &
\overline{1069} & \overline{947} & \overline{591}
& {1219} & \overline{142} & \overline{149} \\
\overline{155} & {240} & {241} & \overline{209} & \overline{91} &
\overline{485} & {169} & \overline{108} & {328} & {107} & {9} &
{211} & {61} & {2823} & \overline{2980}
& {709} & {766} \\
{25} & {104} & {461} & {235} & {185} & {385} & {19} & {164} &
\overline{12} & {13} & \overline{105} & \overline{196} & {61} &
\overline{957} & \overline{543} & \overline{16}
& \overline{18} \\
\overline{231} & {70} & {131} & {1092} & \overline{356} &
\overline{431} & \overline{14} & {163} & \overline{96} &
\overline{11} & \overline{978} & {1039} & \overline{912}
& \overline{763} & \overline{383} & \overline{219} & \overline{269} \\
\overline{18} & {60} & {52} & \overline{1045} & \overline{25} &
\overline{226} & \overline{19} & \overline{20} & \overline{30} &
{215} & \overline{1069} & {1008}
& {999} & \overline{289} & {68} & {43} & {58} \\
\overline{429} & {251} & \overline{297} & {144} & {28} & {12} &
\overline{11} & {117} & {20} & \overline{189} & {73} &
\overline{348} & \overline{3} & {869} & \overline{296}
& \overline{116} & \overline{27} \\
{188} & {108} & \overline{252} & \overline{157} & {25} &
\overline{151} & \overline{176} & \overline{26} & \overline{29} &
\overline{129} & {57} & {31} & \overline{143} &
\overline{702} & \overline{576} & \overline{81} & \overline{94} \\
\overline{36} & \overline{205} & \overline{318} & \overline{1202}
& {182} & {121} & {412} & \overline{284} & {247} & {40} & {1032} &
\overline{1059} & {951} & \overline{726}
& \overline{142} & {498} & {545} \\
\overline{62} & {13} & \overline{11} & {540} & {884} & {2342} &
{2} & {27} & \overline{2019} & \overline{73} & \overline{118} &
{46} & \overline{73} & \overline{706} & {341}
& \overline{1979} & \overline{1883} \\
{1} & \overline{0} & {14} & {957} & {2060} & {393} &
\overline{151} & {7} & \overline{4} & \overline{2158} &
\overline{978} & \overline{949} & {799} & {918} & \overline{634}
& {52} & \overline{56} \\
\overline{2660} & {2593} & {288} & {153} & {8} & {53} &
\overline{28} & \overline{302} & \overline{58} & \overline{4} &
{21} & \overline{68} & {176} & {821} & \overline{247}
& \overline{127} & \overline{123} \\
{48} & \overline{10} & \overline{2468} & \overline{1111} & {38} &
{93} & \overline{11} & {2501} & \overline{18} & \overline{42} &
{1153} & {1010} & \overline{1172} & \overline{482}
& {390} & {6} & \overline{85} \\
\end{array}
\right]
$$
The negative elements of $\bar b_{O}$ are marked with an overline.
The rows of the observation vector $\bar{b}_O$ correspond to the
following consecutive measurements: $\oprod{HH}$, $\oprod{HV}$,
$\oprod{VV}$, $\oprod{HD}-\oprod{HA}$, $\oprod{HL}-\oprod{HR}$,
$\oprod{DH}-\oprod{AH}$, $\oprod{LH}-\oprod{RH}$,
$\oprod{VD}-\oprod{VA}$, $\oprod{VL}-\oprod{VR}$,
$\oprod{DV}-\oprod{AV}$, $\oprod{LV}-\oprod{RV}$,
$\oprod{\Psi^+}-\oprod{\Psi^-}$,
$\oprod{\bar\Psi^+}-\oprod{\bar\Psi^-}$,
$\oprod{\Phi^+}-\oprod{\Phi^-}$,
$\oprod{\bar\Phi^+}-\oprod{\bar\Phi^-}$, where $\ket{\bar
\Phi^\pm}=(\ket{HH}\pm i \ket{VV})/\sqrt{2}$ and  $\ket{\bar
\Psi^\pm}=(\ket{HV}\pm i \ket{VH})/\sqrt{2}$.

\subsection{Pauli matrices based tomography}

$$
\bar b_{P} = \left[\begin{array}{ccccccccccccccccc}
 \overline{1277} &  1264 &  132 &  \overline{57} &  133 &  1144 &  40 &  \overline{113} &  \overline{1037} &  95 &  \overline{52} &  170 &  55 &  175 &  \overline{212} &  \overline{1244} &  \overline{1102} \\
 268 &  \overline{197} &  \overline{1241} &  \overline{1129} &  \overline{1115} &  \overline{126} &  134 &  1222 &  \overline{68} &  1100 &  1032 &  984 &  \overline{986} &  \overline{825} &  577 &  \overline{63} &  \overline{59} \\
 \overline{82} &  \overline{2} &  357 &  196 &  80 &  269 &  98 &  95 &  9 &  71 &  \overline{81} &  \overline{114} &  102 &  \overline{128} &  17 &  33 &  38 \\
 107 &  107 &  105 &  39 &  105 &  117 &  \overline{79} &  69 &  \overline{21} &  \overline{58} &  \overline{24} &  \overline{83} &  \overline{41} &  \overline{830} &  \overline{560} &  \overline{49} &  \overline{57} \\
 143 &  \overline{79} &  \overline{1222} &  \overline{216} &  1107 &  \overline{46} &  \overline{132} &  1182 &  242 &  \overline{1070} &  75 &  \overline{65} &  \overline{96} &  243 &  \overline{23} &  268 &  244 \\
 1247 &  \overline{1264} &  146 &  264 &  57 &  1167 &  74 &  \overline{127} &  \overline{940} &  137 &  \overline{74} &  23 &  \overline{13} &  \overline{759} &  248 &  \overline{1145} &  \overline{1002} \\
 \overline{98} &  138 &  225 &  1148 &  \overline{269} &  \overline{276} &  \overline{214} &  224 &  \overline{172} &  \overline{26} &  \overline{1005} &  1049 &  \overline{932} &  \overline{19} &  \overline{120} &  \overline{359} &  \overline{407} \\
 \overline{134} &  \overline{68} &  \overline{94} &  \overline{55} &  \overline{87} &  \overline{155} &  199 &  \overline{61} &  76 &  15 &  27 &  \overline{10} &  20 &  \overline{745} &  \overline{263} &  140 &  138 \\
 \overline{45} &  22 &  196 &  1186 &  \overline{224} &  245 &  121 &  \overline{155} &  \overline{34} &  \overline{14} &  1054 &  \overline{1039} &  \overline{973} &  \overline{151} &  576 &  \overline{93} &  \overline{104} \\
 137 &  \overline{6} &  269 &  \overline{177} &  \overline{60} &  \overline{249} &  90 &  \overline{113} &  154 &  148 &  \overline{32} &  280 &  33 &  977 &  \overline{1342} &  413 &  397 \\
 1319 &  1305 &  1273 &  310 &  \overline{1150} &  \overline{1199} &  \overline{970} &  1284 &  \overline{1100} &  \overline{1130} &  13 &  \overline{27} &  55 &  \overline{34} &  255 &  \overline{1302} &  \overline{1316} \\
 \overline{19} &  \overline{39} &  118 &  90 &  72 &  \overline{35} &  976 &  183 &  174 &  44 &  \overline{20} &  \overline{16} &  \overline{10} &  926 &  1487 &  415 &  724 \\
 \overline{64} &  82 &  248 &  140 &  \overline{249} &  19 &  102 &  \overline{176} &  \overline{65} &  202 &  \overline{16} &  \overline{31} &  26 &  \overline{440} &  644 &  \overline{50} &  \overline{45} \\
 \overline{292} &  246 &  \overline{28} &  \overline{33} &  \overline{32} &  \overline{237} &  79 &  4 &  174 &  \overline{41} &  41 &  \overline{69} &  29 &  1846 &  \overline{1638} &  296 &  370 \\
 29 &  \overline{47} &  91 &  \overline{242} &  \overline{16} &  113 &  \overline{981} &  158 &  \overline{158} &  \overline{44} &  188 &  \overline{200} &  182 &  54 &  274 &  \overline{367} &  \overline{680} \\
 1397 &  1356 &  1362 &  1290 &  1220 &  1313 &  1000 &  1331 &  1124 &  1157 &  1083 &  1052 &  1004 &  1816 &  1815 &  1367 &  1383 \\
\end{array}\right]
$$
The negative elements of $\bar b_{P}$ are marked with an overline.
The rows of the observation vector $\bar{b}_P$ correspond to the
following consecutive measurements: $(\oprod{DD}+\oprod{AA}) -
(\oprod{DA}+\oprod{AD})$, $(\oprod{DL}+\oprod{AR}) -
(\oprod{DR}+\oprod{AL})$, $(\oprod{DH}+\oprod{AV}) -
(\oprod{DV}+\oprod{AH})$, $(\oprod{DH}+\oprod{AH}) -
(\oprod{DV}+\oprod{AV})$, $(\oprod{LD}+\oprod{RA}) -
(\oprod{LA}+\oprod{RD})$, $(\oprod{LL}+\oprod{RR}) -
(\oprod{LR}+\oprod{RL})$, $(\oprod{LH}+\oprod{RV}) -
(\oprod{LV}+\oprod{RH})$, $(\oprod{LH}+\oprod{RH}) -
(\oprod{LV}+\oprod{RV})$, $(\oprod{HD}+\oprod{VA}) -
(\oprod{HA}+\oprod{VD})$, $(\oprod{HL}+\oprod{VR}) -
(\oprod{HR}+\oprod{VL})$, $(\oprod{HH}+\oprod{VV}) -
(\oprod{HV}+\oprod{VH})$, $(\oprod{HH}+\oprod{VH}) -
(\oprod{HV}+\oprod{VV})$, $(\oprod{HD}-\oprod{HA}) +
(\oprod{VD}-\oprod{VA})$, $(\oprod{HL}-\oprod{HR}) +
(\oprod{VL}-\oprod{VR})$, $(\oprod{HH}-\oprod{HV}) +
(\oprod{VH}-\oprod{VV})$, $(\oprod{HH}+\oprod{HV}) +
(\oprod{VH}+\oprod{VV})$.

\section{Error analysis}
\subsection{Estimated variances}
For all the tomographies the vectors of variances for the 17
measured states are given as matrices
$$
\sigma^2(b) = \bbordermatrix{ &\rho_1 & \rho_2 & \dots &
\rho_{17}\cr
 b_1 & \sigma^2 (b_{1,1})  & \sigma^2 (b_{1,2}) & \ldots &\sigma^2 (b_{1,17})\cr
 b_2 & \sigma^2 (b_{2,1}) & \sigma^2 (b_{2,2}) & \ldots & \sigma^2 (b_{2,17})\cr
 \vdots & \vdots   & \vdots  & \ddots & \vdots \cr
 b_N & \sigma^2 (b_{N,1})& \sigma^2 (b_{N,2}) & \ldots  & \sigma^2 (b_{N,17}) }.
$$
They can be approximated directly with  $\sigma^2(b)\approx
\sigma^2(\bar b)=\bar b$ for all the tomographies except the
optimal one and the tomography based on the Pauli matrices.  For
the optimal tomography the matrix of variances reads
$$
\sigma^2(b_{O}) = \left[\begin{array}{ccccccccccccccccc} 2727 &
2575 &  2844 &  1448 &  127 &  193 &  25 &  2955 &  40 &  27 &
1264 &  809
&  1231 &  2762 &  3831 &  113 &  112 \\
 30 &  58 &  116 &  1312 &  2457 &  2364 &  3928 &  72 &  2555 &  2375 &  863 &
1263 &  757 &  2722 &  2773 &  3452 &  4102 \\
 126 &  43 &  61 &  648 &  2282 &  2660 &  13 &  23 &  1892 &  2199 &  1277 &  895
&  1141 &  977 &  347 &  1887 &  1295 \\
 2706 &  2747 &  2426 &  1751 &  14 &  36 &  35 &  2274 &  8 &  26 &  928 &  1240
&  888 &  802 &  310 &  17 &  23 \\
 1352 &  1297 &  1445 &  1362 &  1299 &  1219 &  1936 &  1427 &  1250 &  1198 &
1088 &  1109 &  997 &  2842 &  3309 &  1900 &  2161 \\
 1425 &  1456 &  1624 &  1433 &  1211 &  1207 &  1933 &  1515 &  1104 &  1102 &
1141 &  989 &  995 &  2844 &  3246 &  1700 &  2016 \\
 1291 &  1199 &  1375 &  1015 &  1199 &  1395 &  23 &  1294 &  964 &  1158 &  1252
&  906 &  1190 &  1858 &  2177 &  955 &  669 \\
 1460 &  1427 &  1454 &  1184 &  1068 &  1203 &  16 &  1454 &  785 &  990 &  1058
&  1103 &  973 &  1936 &  2374 &  820 &  583 \\
 1421 &  1388 &  1196 &  1158 &  1076 &  1321 &  29 &  1154 &  981 &  1184 &  1104
&  1071 &  1035 &  935 &  284 &  930 &  660 \\
 1369 &  1359 &  1204 &  1094 &  1225 &  1410 &  22 &  1118 &  998 &  1258 &  1026
&  1104 &  1070 &  900 &  305 &  968 &  691 \\
 1419 &  1428 &  1233 &  1436 &  1124 &  1071 &  1770 &  1182 &  1236 &  1194 &
934 &  1236 &  834 &  1787 &  1563 &  1834 &  2080 \\
 1295 &  1253 &  1196 &  1360 &  1352 &  1269 &  2108 &  1088 &  1328 &  1307 &
1055 &  1125 &  1037 &  1820 &  1674 &  1979 &  2329 \\
 94 &  41 &  98 &  1018 &  2609 &  2596 &  1889 &  60 &  2194 &  2195 &  1209 &
821 &  1075 &  1533 &  1195 &  2565 &  2506 \\
 44 &  53 &  146 &  1065 &  2112 &  2196 &  1772 &  86 &  2074 &  2276 &  1031 &
 1114 &  862 &  1685 &  1628 &  2274 &  2316 \\
 2757 &  2691 &  2431 &  1027 &  127 &  80 &  104 &  2640 &  74 &  27 &  1301 &
1026 &  1262 &  1778 &  2115 &  150 &  144 \\
 2806 &  2583 &  2521 &  1249 &  160 &  164 &  105 &  2580 &  55 &  67 &  1183 &
 1049 &  1227 &  1737 &  1964 &  306 &  274 \\
\end{array}
\right].
$$
For the Pauli matrices based tomography the matrix of variances
reads
$$
\sigma^2(b_{P}) = \left[\begin{array}{ccccccccccccccccc}
 1323 &  1313 &  1259 &  1206 &  1165 &  1201 &  903 &  1240 &  1093 &  1163 &  1069 &  1052 &  1054 &  1927 &  1799 &  1380 &  1377 \\
 1326 &  1329 &  1327 &  1185 &  1142 &  1257 &  853 &  1254 &  1022 &  1189 &  1075 &  1036 &  1032 &  1886 &  1877 &  1384 &  1340 \\
 1355 &  1314 &  1304 &  1225 &  1162 &  1234 &  897 &  1238 &  1100 &  1176 &  1093 &  1072 &  1012 &  1823 &  1870 &  1395 &  1375 \\
 1355 &  1314 &  1304 &  1225 &  1162 &  1234 &  897 &  1238 &  1100 &  1176 &  1093 &  1072 &  1012 &  1823 &  1870 &  1395 &  1375 \\
 1392 &  1322 &  1323 &  1232 &  1228 &  1235 &  1089 &  1243 &  1089 &  1126 &  1042 &  1072 &  1016 &  1953 &  1952 &  1406 &  1455 \\
 1370 &  1320 &  1322 &  1230 &  1225 &  1225 &  1032 &  1311 &  1051 &  1090 &  1009 &  1062 &  1000 &  1863 &  1965 &  1396 &  1407 \\
 1378 &  1341 &  1325 &  1273 &  1210 &  1236 &  1062 &  1271 &  1057 &  1149 &  1056 &  1114 &  1005 &  1878 &  2024 &  1400 &  1456 \\
 1378 &  1341 &  1325 &  1273 &  1210 &  1236 &  1062 &  1271 &  1057 &  1149 &  1056 &  1114 &  1005 &  1878 &  2024 &  1400 &  1456 \\
 1387 &  1343 &  1321 &  1261 &  1188 &  1270 &  983 &  1291 &  1116 &  1191 &  1097 &  1090 &  1016 &  1889 &  1797 &  1416 &  1411 \\
 1397 &  1408 &  1414 &  1264 &  1218 &  1309 &  978 &  1317 &  1051 &  1180 &  1084 &  1047 &  1033 &  1872 &  1776 &  1334 &  1354 \\
 1397 &  1356 &  1362 &  1290 &  1220 &  1313 &  1000 &  1331 &  1124 &  1157 &  1083 &  1052 &  1004 &  1816 &  1815 &  1367 &  1383 \\
 1397 &  1356 &  1362 &  1290 &  1220 &  1313 &  1000 &  1331 &  1124 &  1157 &  1083 &  1052 &  1004 &  1816 &  1815 &  1367 &  1383 \\
 1387 &  1343 &  1321 &  1261 &  1188 &  1270 &  983 &  1291 &  1116 &  1191 &  1097 &  1090 &  1016 &  1889 &  1797 &  1416 &  1411 \\
 1397 &  1408 &  1414 &  1264 &  1218 &  1309 &  978 &  1317 &  1051 &  1180 &  1084 &  1047 &  1033 &  1872 &  1776 &  1334 &  1354 \\
 1397 &  1356 &  1362 &  1290 &  1220 &  1313 &  1000 &  1331 &  1124 &  1157 &  1083 &  1052 &  1004 &  1816 &  1815 &  1367 &  1383 \\
 1397 &  1356 &  1362 &  1290 &  1220 &  1313 &  1000 &  1331 &  1124 &  1157 &  1083 &  1052 &  1004 &  1816 &  1815 &  1367 &  1383 \\
\end{array}
\right].
$$
\subsection{Estimated error radii}
For each tomography and reconstructed state we have estimated the
maximum error $R_{{}}$ as described in the main text. Our results
are summarized in the following matrix:
$$
R = \bbordermatrix{ &\text{Optimal} &  \text{ MUB} &
\text{Standard}
 & \text{Pauli} & \text{ JKMW}\cr
\rho_1 & 0.0983 &   0.1475 &   0.2183  & 0.2407 &   0.5712 \cr
\rho_2 &  0.0997 &   0.1554 &   0.2221  & 0.2448 &   0.5357\cr
\rho_3 &  0.0987 &   0.1506 &   0.2231  & 0.2422 &   0.5297\cr
\rho_4 &  0.0997 &   0.1477 &   0.2130  & 0.2304 &   0.5044\cr
\rho_5 &  0.1051 &   0.1622 &   0.2344  & 0.2566 &   0.5767 \cr
\rho_6 &  0.0999 &   0.1589 &   0.2291  & 0.2449 &   0.5553\cr
\rho_7 &  0.1152 &   0.2059 &   0.2637  & 0.3222 &   0.6065\cr
\rho_8 &  0.0999 &   0.1580 &   0.2299  & 0.2455 &   0.5850\cr
\rho_9 &  0.1083 &   0.1712 &   0.2507  & 0.2682 &   0.6135\cr
\rho_{10} &  0.1083 &   0.1633 &   0.2388 & 0.2668 &   0.5807\cr
\rho_{11} &  0.1130 &   0.1526 &   0.2281 & 0.2511 &   0.5829\cr
\rho_{12} &  0.1135 &   0.1597 &   0.2293 & 0.2563 &   0.6009\cr
\rho_{13} &  0.1179 &   0.1570 &   0.2318 & 0.2631 &   0.5850\cr
\rho_{14} &  0.0861 &   0.1240 &   0.1509 & 0.2044 &   0.3907\cr
\rho_{15} &  0.0864 &   0.1281 &   0.1587 & 0.2128 &   0.4412\cr
\rho_{16} &  0.0994 &   0.1522 &   0.2171 & 0.2479 &   0.5441\cr
\rho_{17} &  0.0985 &   0.1562 &   0.2168 & 0.2531 &   0.5229\cr
}.
$$
Note that the values are multiplied by a factor of $1.3$ to
compensate for underestimation of $\|\sigma\|$. Standard error is
simply given by $r=R/2$.

\subsection{Relative trace distances between the reconstructed states}
In order to compare the quality of the  matrices reconstructed
with different tomographic protocols, we have also calculated the
relative trace distances for the respective states in each
protocols. Here we omitted the  JKMW protocol, because it
provides the largest error radius. Having three relative distances
(for the remaining three protocols) it is possible to visualize
the relative distances between the matrices and their error radii
on a plane.

$$T =
\bbordermatrix{ & T(\rho_O,\rho_M) &   T(\rho_O,\rho_S) &
T(\rho_M,\rho_S)  \cr
 \rho_1 & 0.1415 &   0.1004 &   0.1203 \cr
 \rho_2 & 0.1462 &   0.0798 &   0.1048 \cr
 \rho_3 & 0.1018 &   0.1130 &   0.1362 \cr
 \rho_4 & 0.1295 &   0.1786 &   0.1967 \cr
 \rho_5 & 0.0818 &   0.1684 &   0.1924 \cr
 \rho_6 & 0.1234 &   0.1176 &   0.0818 \cr
 \rho_7 & 0.0806 &   0.0483 &   0.0990 \cr
 \rho_8 & 0.1155 &   0.1150 &   0.1541 \cr
 \rho_9 & 0.1613 &   0.1245 &   0.0998 \cr
 \rho_{10} & 0.1397 &   0.0956 &   0.1404 \cr
 \rho_{11} & 0.0590 &   0.0515 &   0.0591 \cr
 \rho_{12} & 0.1000 &   0.1074 &   0.0960 \cr
 \rho_{13} & 0.0896 &   0.1000 &   0.0998 \cr
 \rho_{14} & 0.0688 &   0.0425 &   0.0686 \cr
 \rho_{15} & 0.0718 &   0.0790 &   0.0813 \cr
 \rho_{16} & 0.1506 &   0.1270 &   0.1311 \cr
 \rho_{17} & 0.1576 &   0.1278 &   0.1179 \cr
}$$

\begin{figure}
\includegraphics[width=4.25cm]{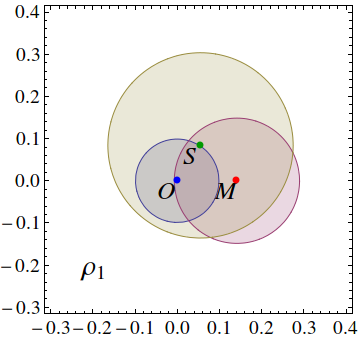}
\includegraphics[width=4.25cm]{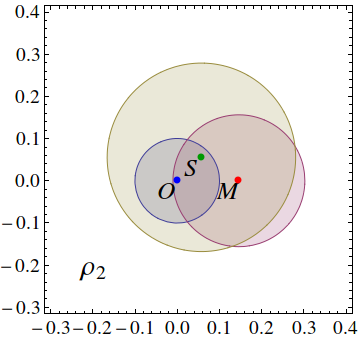}
\includegraphics[width=4.25cm]{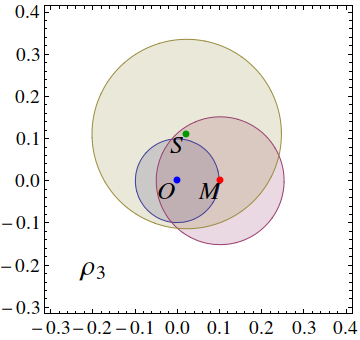}
\includegraphics[width=4.25cm]{rho4}
\includegraphics[width=4.25cm]{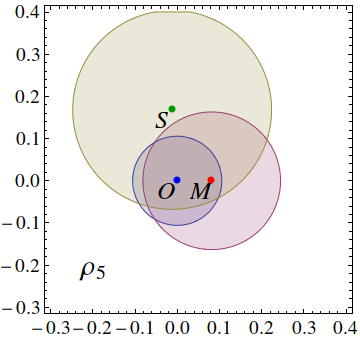}
\includegraphics[width=4.25cm]{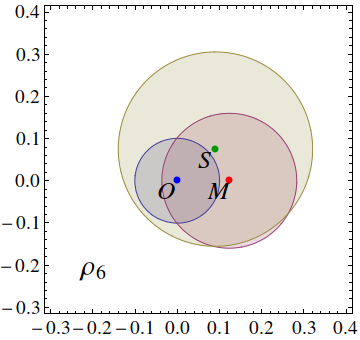}
\includegraphics[width=4.25cm]{rho7}
\includegraphics[width=4.25cm]{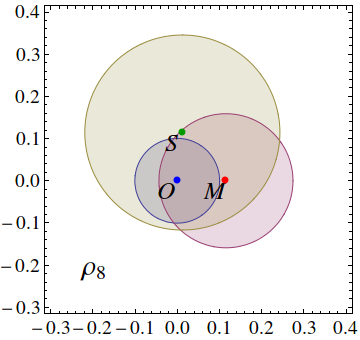}
\includegraphics[width=4.25cm]{rho9}
\includegraphics[width=4.25cm]{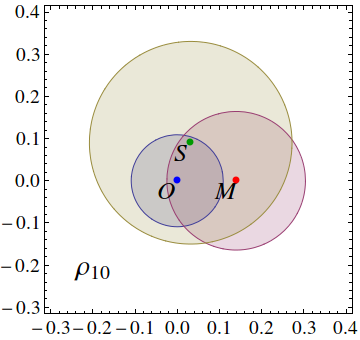}
\includegraphics[width=4.25cm]{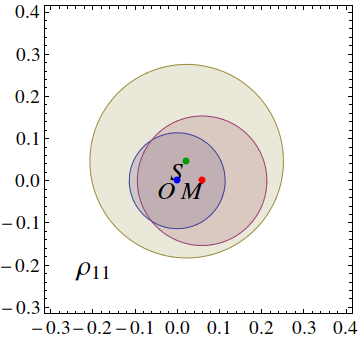}
\includegraphics[width=4.25cm]{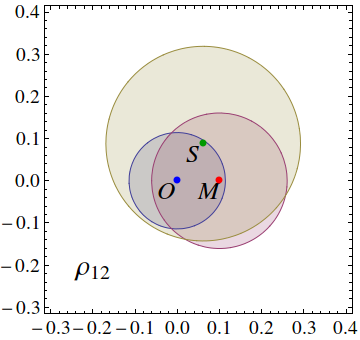}
\includegraphics[width=4.25cm]{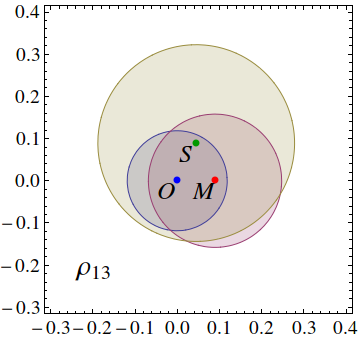}
\includegraphics[width=4.25cm]{rho14}
\includegraphics[width=4.25cm]{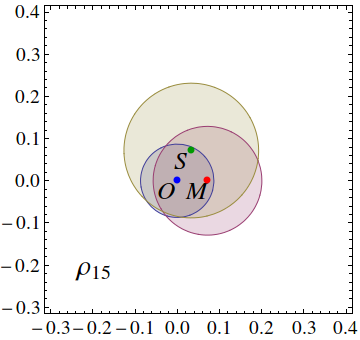}
\includegraphics[width=4.25cm]{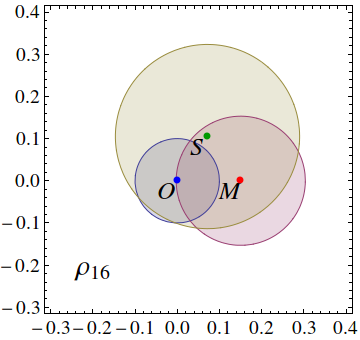}
\includegraphics[width=4.25cm]{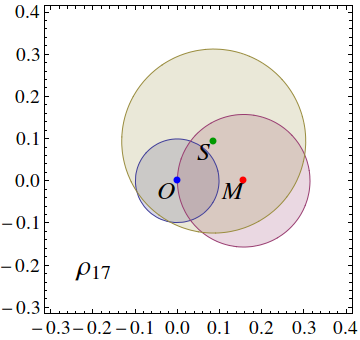}
\caption{Relative distances between points representing the
reconstructed density matrices and their corresponding 
circles of the maximum errors $R_{{}}$ optimal ($O$), standard
($S$), MUB-based ($M$)  tomographies. The 12 states $\rho_n$  are
given in the units of trace distance. The states  can be
approximated by using Eq.~(\ref{eq:states}) with
$\rho_n=\ket{\psi_n}\bra{\psi_n}$. The absolute positions of the
three points are irrelevant. All the states  for $n=1,...,12$ are
fully entangled except $\rho_7$. The states of for $n=13,...,17$
are partially entangled or separable ($n=13,\,14$). The  ideally
reconstructed state  lies in the intersection of the  error 
circles of radius $R_{{}}$.}
\end{figure}
\end{widetext}
\end{document}